\documentclass[12pt]{iopart}

\usepackage{graphicx}
\usepackage{iopams}
\usepackage{bm}
\usepackage{cite}
\usepackage{dsfont}

\usepackage[linktocpage=true]{hyperref}
\hypersetup{
    colorlinks,
    citecolor=blue,
    filecolor=blue,
    linkcolor=blue,
    urlcolor=blue
}

\newcommand{\beg}{\begin{equation}}
\newcommand{\en}{\end{equation}}
\newcommand{\begs}{\begin{subequations}}
\newcommand{\ens}{\end{subequations}}
\newcommand \bea {\begin{eqnarray}}
\newcommand \eea {\end{eqnarray}}
\newcommand{\bem}{\begin{bmatrix}}
\newcommand{\enm}{\end{bmatrix}}
\newcommand{\bpm}{\begin{pmatrix}}
\newcommand{\epm}{\end{pmatrix}}
\newcommand{\bvm}{\begin{vmatrix}}
\newcommand{\evm}{\end{vmatrix}}
\newcommand{\ba}{\begin{array}}
\newcommand{\ea}{\end{array}}

\newcommand{\re}[1]{(\ref{#1})}
\newcommand{\Esref}[1]{Equations~(\ref{#1})}

\newcommand{\Fsref}[1]{Figures~\ref{#1}}

\def\zv{z_{1}, \ldots, z_{N}}
\def\qv{q_{1}, \ldots, q_{M}}

\def\rqv{q^{\mathrm{reg}}_{1}, \ldots, q^{\mathrm{reg}}_{M}}
\def\sqv{q^{\infty}_{1}, \ldots, q^{\infty}_{M}}
\def\ksqv{q^{k\infty}_{1}, \ldots, q^{k\infty}_{M}}
\def\nv{n_{1}, \ldots, n_{N}}
\def\wv{w_{1}, \ldots, w_{K}}
\def\nc{\mathcal{C}}

\def\Zv{Z_{1}, \ldots, Z_{M}}
\newcommand{\ket}[1]{|#1\rangle}

\def\mean#1{\langle #1 \rangle}

\newcommand*\Diff[1]{\mathop{}\!\mathrm{d^#1}}

\newcommand{\blue}[1]{{\color{blue}{#1}}}

\begin{document}

\title{Continuum limit of lattice quasielectron wavefunctions}

\author{Aniket Patra$^{1}$,  Birgit Hillebrecht$^{1}$, and Anne E B Nielsen$^{1,2}$}

\address{$^1$ Max-Planck-Institut f\"{u}r Physik komplexer Systeme, D-01187 Dresden, Germany}
\address{$^2$ Department of Physics and Astronomy, Aarhus University, Ny Munkegade 120, DK-8000 Aarhus C, Denmark}

\begin{abstract}

Trial states describing anyonic quasiholes in the Laughlin state were found early on, and it is therefore natural to expect that one should also be able to create anyonic quasielectrons. Nevertheless, the existing trial wavefunctions for quasielectrons show behaviors that are not compatible with the expected topological properties or their construction involves ad hoc elements. It was shown, however, that for lattice fractional quantum Hall systems, it is possible to find a relatively simple quasielectron wavefunction that has all the expected properties [\href{https://iopscience.iop.org/article/10.1088/1367-2630/aab5d0}{New J. Phys.\ \textbf{20}, 033029 (2018)}]. This naturally poses the question: what happens to this wavefunction in the continuum limit? Here we demonstrate that, although one obtains a finite continuum wavefunction when the quasielectron is on top of a lattice site, such a limit of the lattice quasielectron does not exist in general. In particular, if the  quasielectron is put anywhere else than on a lattice site, the lattice wavefunction diverges when the continuum limit is approached. The divergence can be removed by projecting the state on the lowest Landau level, but we find that the projected state does also not have the properties expected for anyonic quasielectrons. We hence conclude that the lattice quasielectron wavefunction does not solve the difficulty of finding trial states for anyonic quasielectrons in the continuum.

\end{abstract}

\begin{indented}\item[]
\end{indented}

\section{Introduction}
\label{Sec: Intro}

The discovery of the fractional quantum Hall effect (FQHE) in heterojunctions in solids \cite{fqheorigexpt} marks the genesis of the enduring interest in strongly correlated topologically ordered systems \cite{tong, stern, jainbook}. The topology in the FQHE is manifested in the  presence of quasiholes and quasielectrons, which are anyonic excitations with positive and negative charge, respectively. In such solid-state systems \cite{fqheorigexpt}, similar experimental manipulations --   increasing or decreasing the external magnetic field -- lead to realizations of quasiholes or quasielectrons, respectively.

Kitaev pointed out the usefulness of anyons in fault-tolerant topological quantum computing \cite{kitaev}. The first step in achieving this, by utilizing the anyonic excitations appearing in the FQHE, is to theoretically understand them. The dimension of such systems' Hilbert space being exponential in the number of electrons renders it difficult for obtaining numerical results ab initio. Descriptions in terms of ansatz wavefunctions have traditionally provided the way forward.

For example, Laughlin provided an ansatz wavefunction explaining the FQHE ground state at $1/3$ filling \cite{origlaugh}. In the same paper, he also provided an ansatz for the quasihole excitation, where the electron density is locally decreased, by modifying the ground state wavefunction. However, the difficulty of obtaining an ansatz for the quasielectron,  where the electron density at the location of the quasielectron needs to be increased, was immediately understood. This  leads to a pole singularity due to the Pauli exclusion principle.

Distinct wavefunctions for quasiholes and quasielectrons in the FQHE with different filling fractions were introduced using several different approaches \cite{origlaugh, laughbook, cfqe1, cftqe1, cftqe2, cftqe3, cftqe4, mpsqe,   composite_qe_qh, PBC_op_form, cluster}. Laughlin, in his paper, introduced an ansatz in terms of derivative operators $\partial/\partial Z_{i}$, where $Z_{i}$ are the electron coordinates \cite{origlaugh}. Under the composite fermion framework, Jeon and Jain proposed a different ansatz that has lower variational energy compared to the Laughlin ansatz \cite{cfqe1}. This quasielectron ansatz can be thought of as a composite of a quasiparticle of negative charge $-2/3$ and a quasihole of charge $1/3$, orbiting around the common center of mass with positive relative angular momentum \cite{composite_qe_qh}. The subtleties of the braiding properties of similar wavefunctions for composite fermion quasiparticles were analyzed \cite{cfqe2}. One can obtain a quasielectron wavefunction by using a physical clustering condition, where it must vanish when certain patterns of clusters of electrons are formed \cite{cluster}. Another way to obtain quasielectron and quasihole ans\"{a}tze is to consider correlation functions of certain rational CFTs that involve nonlocal operators \cite{cftqe1, cftqe2, cftqe3}. How to construct all states in the Abelian quantum Hall hierarchy using the above quasihole and quasielectron wavefunctions was also explained \cite{cftqe4}. An operator formalism describing the Laughlin quasielectron in FQH systems with periodic boundary conditions (i.e., on a 2-torus) is also known \cite{PBC_op_form}. In all these descriptions, however, one observes that either the braiding statistics or the positions of the quasielectrons are not as expected for anyonic quasielectrons. A description in which all properties are as expected for anyonic quasielectrons was found in \cite{mpsqe}, but it required some ad hoc modifications of the trial states to obtain this result.

It is possible to write trial wavefunctions for quasiholes and quasielectrons on a lattice \cite{anqe}, which have a natural construction and all the expected topological properties. This description is particularly simple because the quasiparticle operators are written in terms of a single screened chiral boson field, and the density profiles are undistorted due to absence of additional screening requirements. Interesting strongly correlated topological phases such as topological Chern insulators are only present in such lattice systems \cite{FCI1, FCI2, FCI3}. These are, e.g., relevant for implementations in optical lattices. In this paper, we investigate the continuum limit of these lattice wavefunctions. This analysis of the continuum limit -- necessary for the description of solid-state anyonic excitations  -- was absent in Ref.~\cite{anqe}. Although placing the quasielectron on a lattice site produces a finite trial wavefunction, our analytical calculations show that the continuum limit of the quasielectron does not exist in general. Additionally, we show that an attempt to cure the wavefunction of this problem by projecting it on the lowest Landau level (LLL) in the continuum limit, as was alluded to in \cite{origlaugh}, does also not give a suitable trial state for anyonic quasielectrons. In particular, the LLL-projected continuum quasielectron ansatz does not have the expected braiding properties.

We recapitulate the Laughlin wavefunctions for the continuum FQH ground state and the quasihole in \Sref{Sec: Laugh_Cont_Wavefunc}, and explain why the quasielectron -- described as an inverse quasihole -- is singular. We recall  in \Sref{Sec: Lattice_Wavefunc} the wavefunctions for the anyonic excitations on a lattice with $N$ lattice points and their parent Hamiltonians. These wavefunctions are not singular for finite $N$.  In \Sref{Sec: Cont_QE_Numerics}, we approach the continuum limit by increasing the number of the lattice points $N$, while keeping the number of electrons and the number of magnetic fluxes in the system constant. We consider different positions of the quasielectron in relation to the lattice sites. In particular, we explain that, when we put the quasielectron on top of a lattice site, it leads to a finite trial wavefunction. In \Sref{Sec: Cont_QE_Analytical_Argument}, we identify the terms that diverge in the continuum limit. We show that the ratio of the contributions from the divergent and regular terms is proportional to $\ln N$. This means that the continuum quasielectron is singular and is not confined to the LLL. The slow nature of this divergence makes it hard for detection in the numerics. We numerically show that indeed only the divergent terms are responsible for bringing about the most amount of change in the physical observables (e.g., excess charge) as we approach the continuum limit.

To remove the singularity we project the quasielectron on the LLL. We study this modified divergence free ansatz in \Sref{Sec: Failure_Div_Free}, and explain how this also does not possess the topological properties expected for an anyonic quasielectron. In \Sref{Sec: Incorr_Braid} we show that the anyon statistics obtained by moving a LLL-projected quasielectron adiabatically around a quasihole is different from the one obtained by moving the quasihole around the same quasielectron. The Berry phase obtained by moving the LLL-projected quasielectron around  a quasihole depends on the size and shape of the contour. Moreover, in \Sref{Sec: Isolated_QH}, we demonstrate that, if  starting from the ansatz wavefunction with a quasihole and a LLL-projected quasielectron we take the position of the quasielectron to infinity, we fail to retrieve the wavefunction for an isolated quasihole.

\section{Laughlin Wavefunction: Continuum Description}
\label{Sec: Laugh_Cont_Wavefunc}

Consider a FQH droplet of filling fraction $1/q$ with $M$ electrons, where $q$ is a positive integer. To describe $K$ quasiparticles at $\wv$, Laughlin introduced the following ansatz wavefunctions:
\bea
\fl \big |\psi\big \rangle_{\mathrm{cont}} = \nc^{-1}\int\cdots\int\prod_{i = 1}^{M}\Diff2 Z_{i} \big|\Zv\big\rangle \nonumber \\
\times \prod_{i, k}\left(w_{k} - Z_{i}\right)^{p_{k}}  \prod_{i<j} \left(Z_{i} - Z_{j}\right)^{q} \rme^{-\sum_{i = 1}^{M}|Z_{i}|^2/4l_{B}^2},
\label{Laugh_QP}
\eea
where $\nc$ is a real normalization constant, the electronic coordinates $Z_{i}$ take any value on the complex plane, and $l_{B}$ is the magnetic length \cite{origlaugh}. The Jastrow factor prevents two particles from being at the same position. The state describes fermions for $q$ odd and hardcore bosons for $q$ even.

There are two types of quasiparticles in the FQHE, namely the quasihole and the quasielectron with fractional charges $+e/q$ and $-e/q$, respectively. Here $e>0$ is the absolute value of the electronic charge. In \Eref{Laugh_QP}, $p_{k} = 0$ for all $k$, produces the Laughlin FQHE ground state wavefunction. When $p_{k} = +1$, there is a quasihole at $w_{k}$, where the additional (local in $w_{k}$) factor $\prod_{i}\left(w_{k} - Z_{i}\right)$ corresponds to the insertion of a positive flux tube at $w_{k}$ \cite{origlaugh}.

However, the wavefunction is not physical when $p_{k} = -1$. The factor corresponding to the insertion of a negative flux tube at $w_{k}$ is singular, when any of the electronic coordinates $Z_{i}$ are equal to $w_{k}$. This renders the naive description of the quasielectron to be an inverse quasihole invalid.

\section{Lattice Wavefunctions and Hamiltonians}
\label{Sec: Lattice_Wavefunc}

In this section, we recall the salient features of the quasielectron wavefunction on a lattice, which was introduced in Ref.~\cite{anqe}. The lattice sites are $z_{i}$ for $i = 1, \ldots, N$. Here we consider the case of a square lattice with a roughly circular boundary of radius $R$. The local basis on site $i$ is labelled by $\ket{n_{i}}$, where the lattice occupancy $n_{i}$ is either $0$ or $1$. After setting the magnetic length $l_{B} = 1$, it is possible to write down the wavefunction for $K$ different quasiparticles at the positions $\wv$ in terms of the chiral correlator of CFT vertex operators as follows:
\beg
\big |\psi \big \rangle \propto \sum_{\nv} \langle 0| \prod_{k = 1}^{K}W_{p_{k}} \prod_{i = 1}^{N}V_{n_{i}} |0\rangle \big |\nv \big \rangle,
\label{Latt_QP_CFT_Corr}
\en
where the vertex operators are
\beg
\eqalign{W_{p_{k}} = {} :\rme^{\rmi p_{k}\phi\left(w_{j}\right)/\sqrt{q}}:, \\
V_{n_{j}} = \chi_{n_{j}}:\rme^{\rmi \left(qn_{j} - \eta\right)\phi\left(z_{j}\right)/\sqrt{q}}:.}
\label{CFT_Vertex}
\en
Here $:\ldots :$ denotes normal ordering, $\phi\left(z_{j}\right)$ is the chiral field of a massless free boson evaluated at the position $z_{j}$, $\chi_{n_{j}}$ are the unspecified single particle phase factors that do not depend on the $w_{j}$, and the area assigned per single lattice site is $2\pi\eta$. The vertex operator $W_{p_{k}}$ with $p_{k} = +1\;(-1)$ creates a quasihole (quasielectron).

From \Eref{Latt_QP_CFT_Corr}, we obtain the following ansatz for the $K$ quasiparticle excitations at the positions~$\wv$:
\beg
\fl \big |\psi \big \rangle = \mathcal{C}^{-1} \sum_{\nv}\delta_{n}\prod_{i = 1}^{N}\chi_{n_{i}} \prod_{i, k}\left(w_{k} - z_{i}\right)^{p_{k}n_{i}} \prod_{i<j}\left(z_{i} - z_{j}\right)^{qn_{i}n_{j} - \eta\left(n_{i} + n_{j}\right)} \big |\nv \big \rangle,
\label{Latt_QP}
\en
where the unspecified single particle phase factors are written as
\beg \label{SP_Phase}
\chi_{n_{i}} = \rme^{\rmi\left(\phi_{0} + \phi_{j}n_{j}\right)},\quad \phi_{0}, \phi_{j} \in \mathbb{R}.
\en
Note that the correlation function is zero unless the charge neutrality condition,
\beg
\sum_{i = 1}^{N} n_{i} = \left(\eta N - \sum_{k = 1}^{K} p_{k}\right)/q,
\label{Charge_Neutrality}
\en
is obeyed by the lattice occupancies. This is imposed in \Eref{Latt_QP} by the Kronecker delta function $\delta_{n}$. The real normalization constant $\mathcal{C}$ in \Eref{Latt_QP} is a function of the quasiparticle positions, and it obeys
\beg
\fl \mathcal{C}^{2} = \sum_{\nv}\delta_{n}  \prod_{k,j}\left|w_{k} - z_{j}\right|^{2p_{k}n_{j}} \prod_{i<j}\left|z_{i} - z_{j}\right|^{2qn_{i}n_{j} - 2\eta\left(n_{i} + n_{j}\right)} = \sum_{\nv} \delta_{n}\nc_{\nv}^{2}.
\label{Norm_Sq}
\en

\Eref{Latt_QP} provides a valid (finite) trial wavefunction. This ansatz also works when the quasielectron position $w_{k}$ is arbitrarily close to a lattice site $z_{i}$. In this case, the proximity of $w_{k}$ and $z_{i}$ forces $n_{i} \rightarrow 1$. This allows us to include the infinitely large constant $(w_{k} - z_{i})^{-1}$ in the normalization.

The charge neutrality condition described in \Eref{Charge_Neutrality} guarantees that the number of particles in the FQH droplet remains constant for any $\eta$. This in turn means $N\eta =\mathrm{constant,}$ where $N$ is the number of lattice sites inside the circle of fixed radius $R$. As we decrease $\eta$, $R$ stays fixed, but the number of sites inside the circle increases, which allows the particles to be in more places. Thus the parameter $\eta$ allows us to interpolate between the lattice limit ($\eta = 1$) and the continuum limit ($\eta \rightarrow 0^{+}$). In the continuum limit and with $p_{k} = 0$ for all $k$, starting from the wavefunction \re{Latt_QP} defined on a lattice with uniformly distributed sites and a circular boundary, one obtains the Laughlin wavefunction. In the continuum limit, the occupied subset of lattice points ($M$ of them) are the electronic coordinates~$\Zv$ \cite{lattlaugh}.

The wavefunction \re{Latt_QP} is the exact ground state of the following few-body Hamiltonian \cite{anqe}:
\bea \label{H}
H &= \sum_{i = 1}^{N}\hat{\Lambda}^{\dagger}_{i}\hat{\Lambda}_{i}, \nonumber \\
\hat{\Lambda}_{i} &= \sum_{j (\neq i)} \frac{1}{z_{i} - z_{j}}\left[T_{j}^{-1}\hat{d}_{j} - T_{i}^{-1}\hat{d}_{i}\left(q\hat{n}_{j} - 1\right)\right], \\
T_{k} &= \rme^{i\phi_{k}}\rme^{-i\pi(k-1)}\prod_{i}\left(w_{i} - z_{k}\right)^{p_{i}}\prod_{j}\left(z_{j} - z_{k}\right)^{1-\eta}, \nonumber
\eea
where $\phi_{k}$ were introduced in \Eref{SP_Phase}, $\hat{d}_{k}$ is the hardcore fermionic (bosonic) annihilation operator acting on site $k$ for q odd (even), and $\hat{n}_{k} = \hat{d}^{\dagger}_{k}\hat{d}_{k}$ is the number operator for the $k$th site. The above \Eref{H} is a valid Hamiltonian as long as
\beg \label{H_ineq}
\eta - \sum_{i}p_{i}/N< 1 + q/N.
\en
This demonstrates that as we take the continuum limit there always exist few-body Hamiltonians, for which the wavefunctions \re{Latt_QP} are exact ground states.

One can also compare the analytical states for $q=2$ to the ground states of an interacting, bosonic Hofstadter model on a square lattice introduced in \cite{FQH_OL}. The model allows nearest neighbor hopping, and the hopping terms are complex to mimic a uniform magnetic field. The on-site interactions are so strong that there is at most one boson on each site. For periodic boundary conditions, it was found in \cite{FQH_OL_PRA} that this model is in a topological phase for $\eta$ less than about $0.4$, and there is a high overlap with the lattice Laughlin states on a torus for $\eta$ less than about $0.3$. The analytical states can be obtained for periodic boundary conditions by evaluating the CFT correlator describing the state on the torus. This gives $q$ states due to the topological degeneracy on the torus. In \cite{anqe} it was found that the analytical states on a $6\times 6$ square lattice on a torus with one quasihole, one quasielectron, $q=2$, and three particles (giving $\eta=1/6$) both have overlaps of $0.99$ with the ground states of the interacting Hofstadter model with a positive and a negative pinning potential added to trap two anyons of opposite charge.

\subsection{Braiding Statistics}
\label{Sec: Braiding_Lattice_Wavefunc}

When the $k$th anyon situated at $w_{k}$ moves adiabatically in a closed contour $c$, the wavefunction is modified as $\big |\psi \big \rangle \rightarrow \mathbb{M}\rme^{i\theta_{k}}\big |\psi \big \rangle$. Here $\theta_{k}$ is the Berry phase, and the monodromy $\mathbb{M}$ is the change in the wavefunction effected by only the analytical continuation.

For future use, we briefly repeat the computation of the Berry phase $\theta_{k}$. Starting with the wavefunction \re{Latt_QP} for $K$ anyons, we write $\theta_{k}$ as follows:
\beg 
\theta_{k} = \rmi\oint_{c}\bigg\langle\psi\bigg |\frac{\partial \psi}{\partial w_{k}}\bigg\rangle\, dw_{k} + \mathrm{c.c.} = \frac{\rmi}{2}\oint_{c}\frac{1}{\nc^{2}}\frac{\partial \nc^{2}}{\partial w_{k}} \, dw_{k} + \mathrm{c.c.}.
\label{BC_Def}
\en%
From \Eref{Norm_Sq} we observe that
\beg
\frac{\partial \nc^{2}}{\partial w_{k}} = \sum_{\nv} \delta_{n} \nc_{\nv}^{2}\sum_{j}\frac{p_{k}n_{j}}{w_{k} - z_{j}}.
\label{BC_deriv_calc}
\en
We also recall the following expression for the expectation value of an operator $\hat{\mathcal{O}}$ that is diagonal in the $|\nv\rangle$ basis:
\beg
\left\langle\psi\left|\hat{\mathcal{O}}\right|\psi\right\rangle
= \frac{1}{\nc^{2}} \sum_{\nv} \delta_{n}\nc_{\nv}^{2} \left\langle \nv \left|\hat{\mathcal{O}}\right| \nv \right\rangle.
\label{Exp_Value}
\en 
Using \Esref{BC_Def}, \re{BC_deriv_calc}, and \re{Exp_Value}, we obtain \cite{braid1, braid2}
\beg
\theta_{k} = i\frac{p_{k}}{2}\oint_{c}\sum_{i}\frac{\mean{n_{i}}}{w_{k} - z_{i}} dw_{k} + \mathrm{c.c.}.
\label{BP_Exp}
\en

When the quasiparticles are screened -- i.e., a quasiparticle at $w_{j}$ only affects the particle densities close to $w_{j}$ (see \Fref{Loc_QE}, for example) -- it is possible to separate them sufficiently. We now move a quasiparticle at $w_{k}$ in a closed contour $c$, such that $w_{j}$ always remains far from $c$. One then writes the anyonic statistics $\gamma$ of the $k$th and the $j$th quasiparticles as
\bea 
\fl 2\pi\gamma = \theta_{k, \left(w_{j} \mathrm{ inside}\right)} - \theta_{k, \left(w_{j} \mathrm{ outside}\right)} \nonumber \\
= \frac{\rmi p_{k}}{2}\oint_{c}\sum_{i}\frac{\mean{n_{i}}_{\left(w_{j} \mathrm{ inside}\right)} - \mean{n_{i}}_{\left(w_{j} \mathrm{ outside}\right)}}{w_{k} - z_{i}} dw_{k} + \mathrm{c.c.},
\label{Anyon_Stat}
\eea
where $\theta_{k, \left(w_{j} \mathrm{ inside}\right)} \left( \theta_{k, \left(w_{j} \mathrm{ outside}\right)}\right)$ is the Berry phase with the $j$th anyon inside (outside) $c$. In the above equation, we have used the fact that the monodromy $\mathbb{M}$ is the identity.

For screened and well-separated anyons, $\mean{n_{i}}_{\left(w_{j} \mathrm{ inside}\right)} - \mean{n_{i}}_{\left(w_{j} \mathrm{ outside}\right)}$ is independent of $w_{k}$. It is non-zero only for lattice positions that are close to the two possible values of $w_{j}$. Using this in \Eref{Anyon_Stat}, one obtains the anyon statistics 
\begin{equation}\label{statistics}
\gamma = p_{j}p_{k}/q,
\end{equation} 
which is the expected statistics for anyons in a model with the topology of the Laughlin state. We note that the above argument neither depends on the nature (whether they are quasiholes or quasielectrons) of the anyons, nor on the value of $\eta$.

\section{Divergence of the Continuum Limit of the Lattice Wavefunction}
\label{Sec: Div_Cont}

For the sake of definiteness, we consider one quasielectron at $w_{1}$ and one quasihole at $w_{2}$. We write the ansatz for this system using \Eref{Latt_QP} as follows:
\beg 
\fl \big|\psi\big\rangle = \mathcal{C}^{-1} \sum_{\nv}\delta_{n} \prod_{i = 1}^{N}\chi_{n_i}\prod_{i = 1}^{N}\left(\frac{w_{2} - z_{i}}{w_{1} - z_{i}}\right)^{n_{i}} \prod_{i<j}\left(z_{i} - z_{j}\right)^{qn_{i}n_{j} - \eta\left(n_{i} + n_{j}\right)} \big |\nv \big \rangle.
\label{Latt_1QE_1QH}
\en 
Note that, since $\sum_{k} p_{k} = 0$ in this system, we still have $M = N\eta/q$ particles.

\subsection{Continuum Limit of the Quasielectron Ansatz: Numerics}
\label{Sec: Cont_QE_Numerics}

In order to isolate and study the properties of the quasielectron, we consider $w_{1}$ to be the origin and $w_{2}$ to be the complex infinity. In the limit $w_{2}\rightarrow \infty$, the factor $\prod_{i = 1}^{N}\left(w_{2} - z_{i}\right)^{n_{i}}$ approaches $\prod_{i = 1}^{N}w_{2}^{n_{i}} = w_{2}^M$, which is a constant (albeit, an infinite one) that can be absorbed in the normalization. This gives the following wavefunction for a single quasielectron:
\beg 
\fl \big|\psi_{\mathrm{QE}} \big\rangle = \mathcal{C}^{-1} \sum_{\nv}\delta_{n} \prod_{i = 1}^{N}\tilde{\chi}_{n_i}\prod_{i = 1}^{N}z_{i}^{-n_{i}}  \prod_{i<j}\left(z_{i} - z_{j}\right)^{qn_{i}n_{j} - \eta\left(n_{i} + n_{j}\right)} \big |\nv \big\rangle,
\label{Latt_1QE}
\en 
where $\tilde{\chi}_{n_i} = (-1)^{n_{i}}\chi_{n_i}$.


\begin{figure}[tbp!]
\begin{indented}\item[]
\includegraphics[trim={0.7cm 7cm 0.7cm 7cm},clip, scale=0.38]{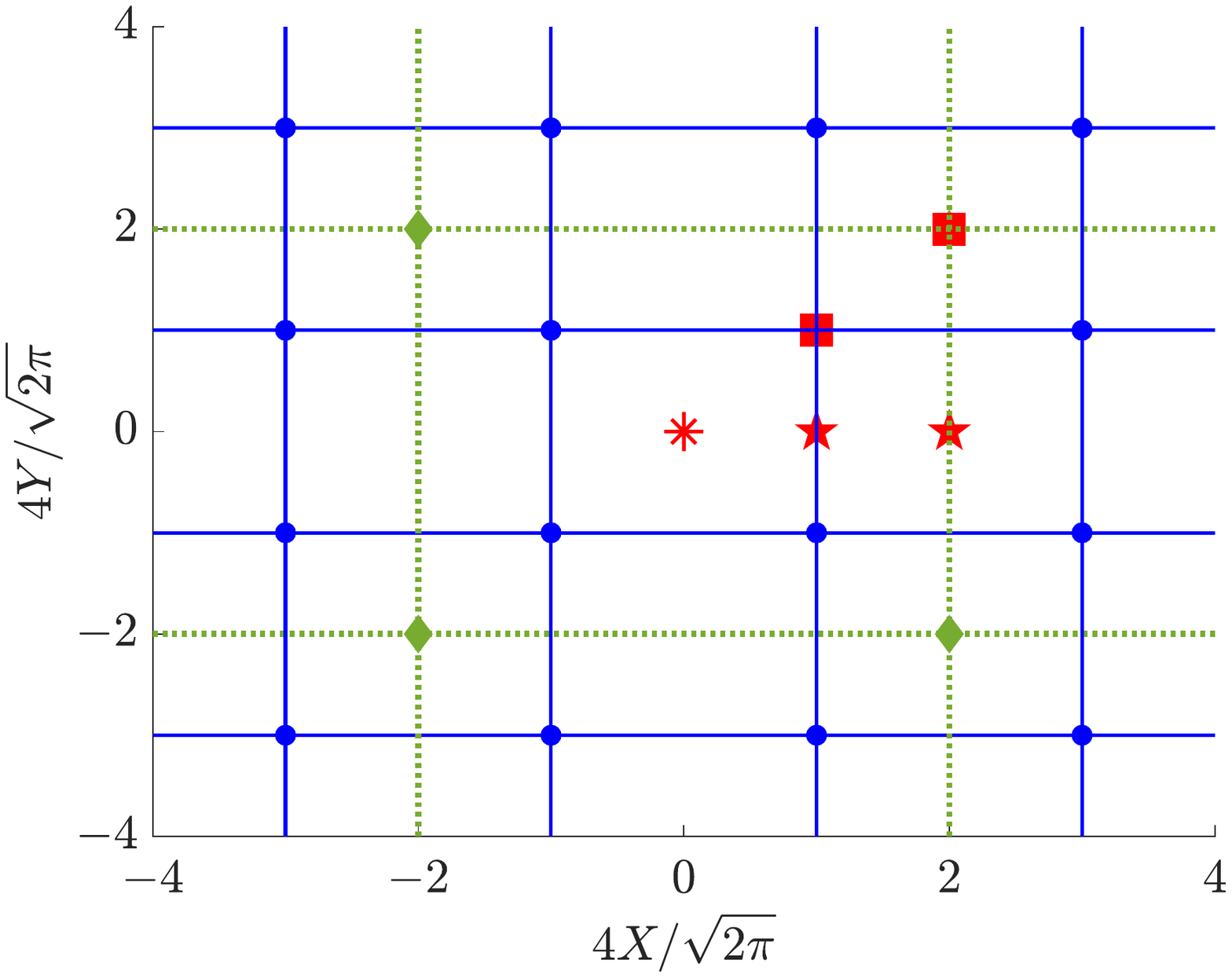}
\caption{A step toward the continuum limit for a square lattice. We start with a square lattice (green diamonds), where the length of the primitive lattice vector is $\sqrt{2\pi}$. In the next step (new lattice points are the blue dots), the length of the primitive lattice vector becomes $\sqrt{\pi/2}$, and we have four times as many lattice sites inside the same area. The three inequivalent quasielectron positions that we consider (cf. \Fref{Diff_Pos}) are -- \textbf{(1)} the center of a square plaquette formed by four lattice points (red asterisk), \textbf{(2)} midway between two lattice sites (red pentagram), and \textbf{(3)} on a lattice point (red square).}
\label{Lattice_Doubling}
\end{indented}
\end{figure}

\begin{figure}[tbp!]
\begin{indented}\item[]
\fl \raisebox{5cm}{\bf(a)} \includegraphics[trim={0.7cm 7cm 0.7cm 7cm},clip, scale=0.38]{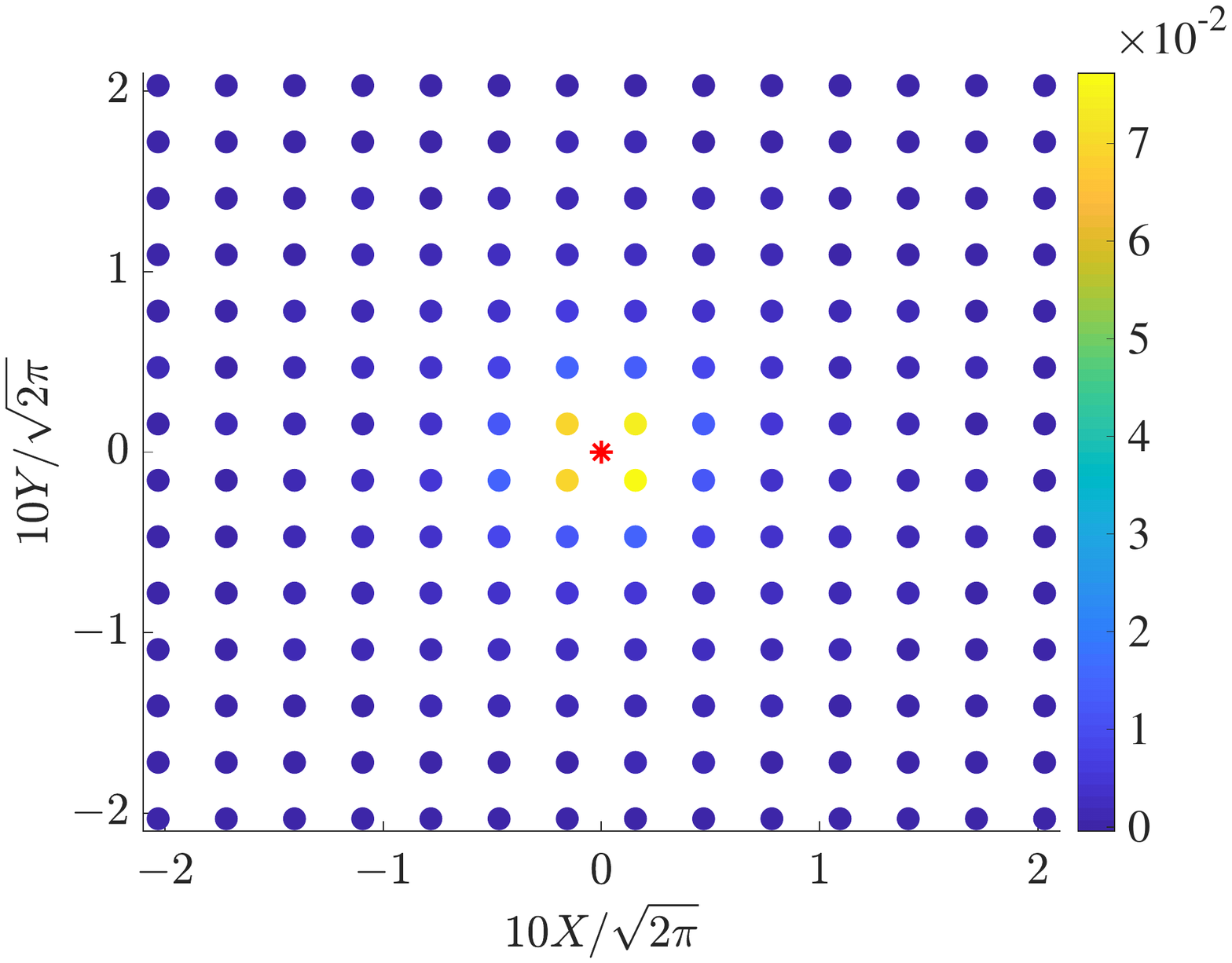} 
\raisebox{5cm}{\bf(b)} \includegraphics[trim={0.7cm 7cm 0.7cm 7cm},clip, scale=0.38]{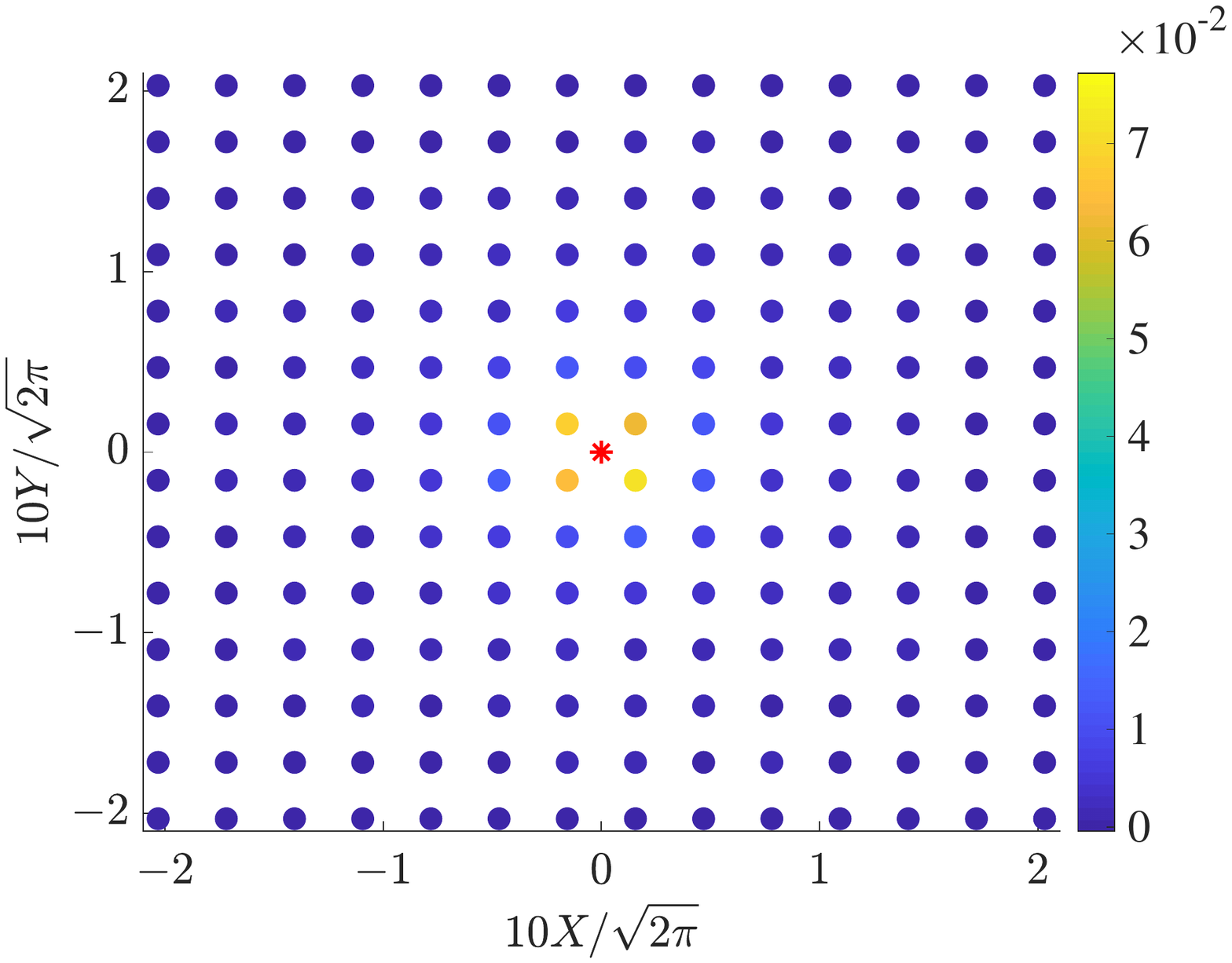} 
\caption{Excess particle densities $\mean{n_{i}}_{-1}- \mean{n_{i}}_{0}$ (see below \Eref{Step_Fn} for the definitions of $\mean{n_{i}}_{-1}$ and $\mean{n_{i}}_{0}$) at different lattice points in the presence of a quasielectron \re{Latt_1QE}, shown by the red star at the origin, for inverse filling fractions $q = 2$ \textbf{(a)}, and $q = 3$ \textbf{(b)}. Here we only show a part of the lattice. The particle densities of only a few lattice sites close to the quasielectron are affected. The screened nature of the quasielectron shows that the quasiparticles described by \Eref{Latt_QP} have the braiding properties \Eref{statistics}, see \Sref{Sec: Braiding_Lattice_Wavefunc}.}
\label{Loc_QE}
\end{indented} 
\end{figure}

\begin{figure}[tbp!]
\begin{indented}\item[]
\fl \raisebox{5cm}{\bf(a)} \includegraphics[trim={0.7cm 7cm 0.7cm 7cm},clip, scale=0.38]{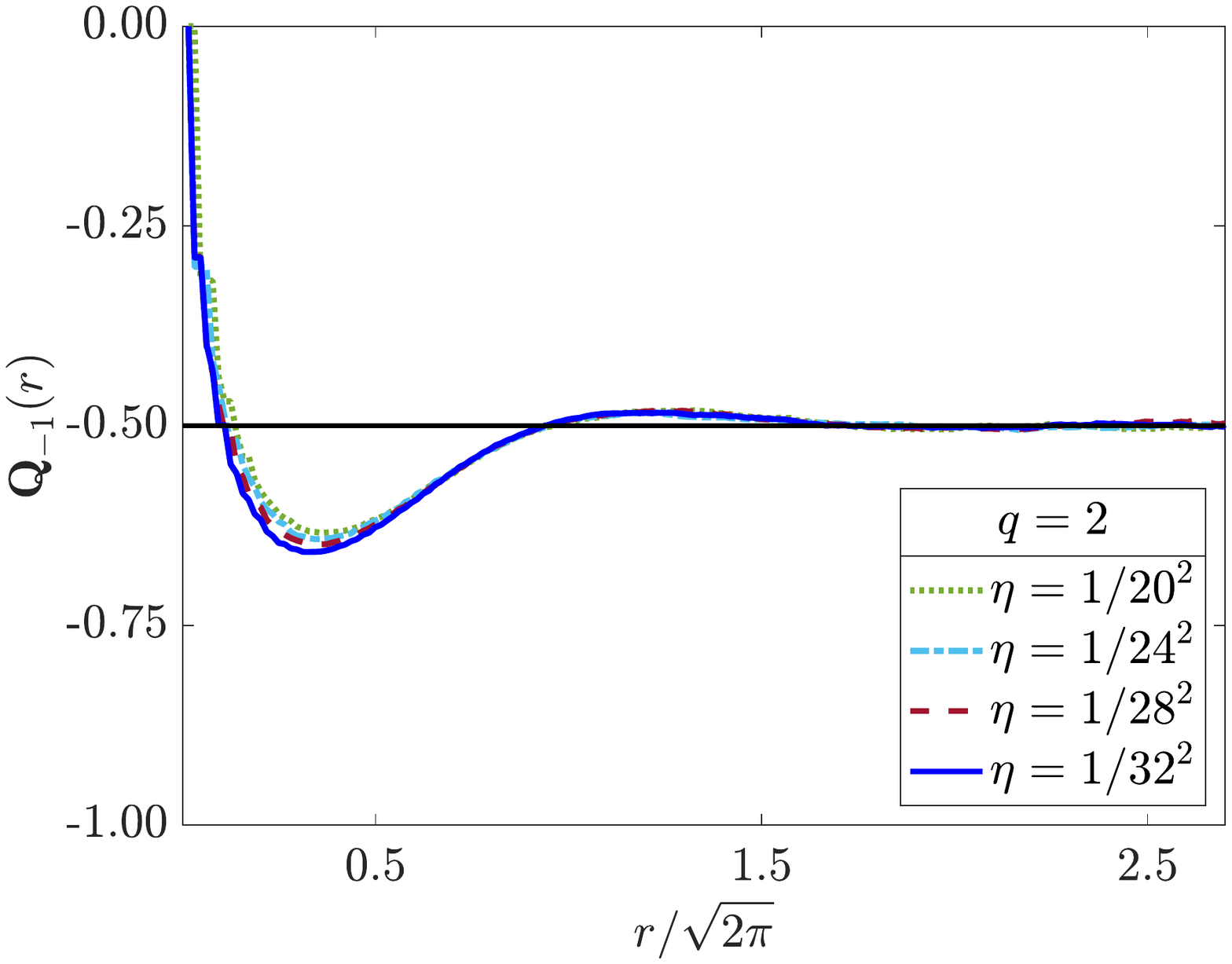}\quad
\raisebox{5cm}{\bf(b)} \includegraphics[trim={0.7cm 7cm 0.7cm 7cm},clip, scale=0.38]{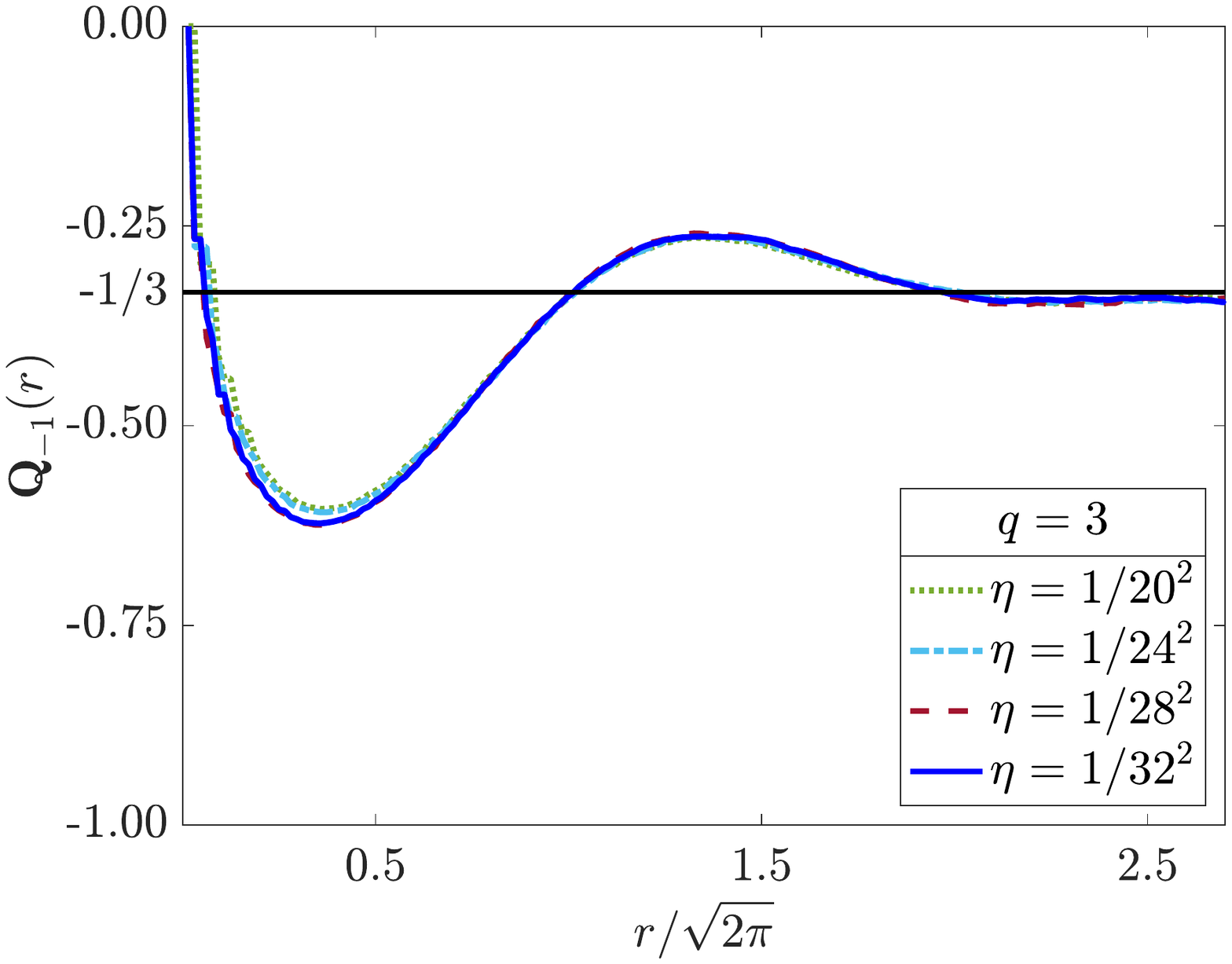} \\
\fl \raisebox{5cm}{\bf(c)} \includegraphics[trim={0.7cm 7cm 0.7cm 7cm},clip, scale=0.38]{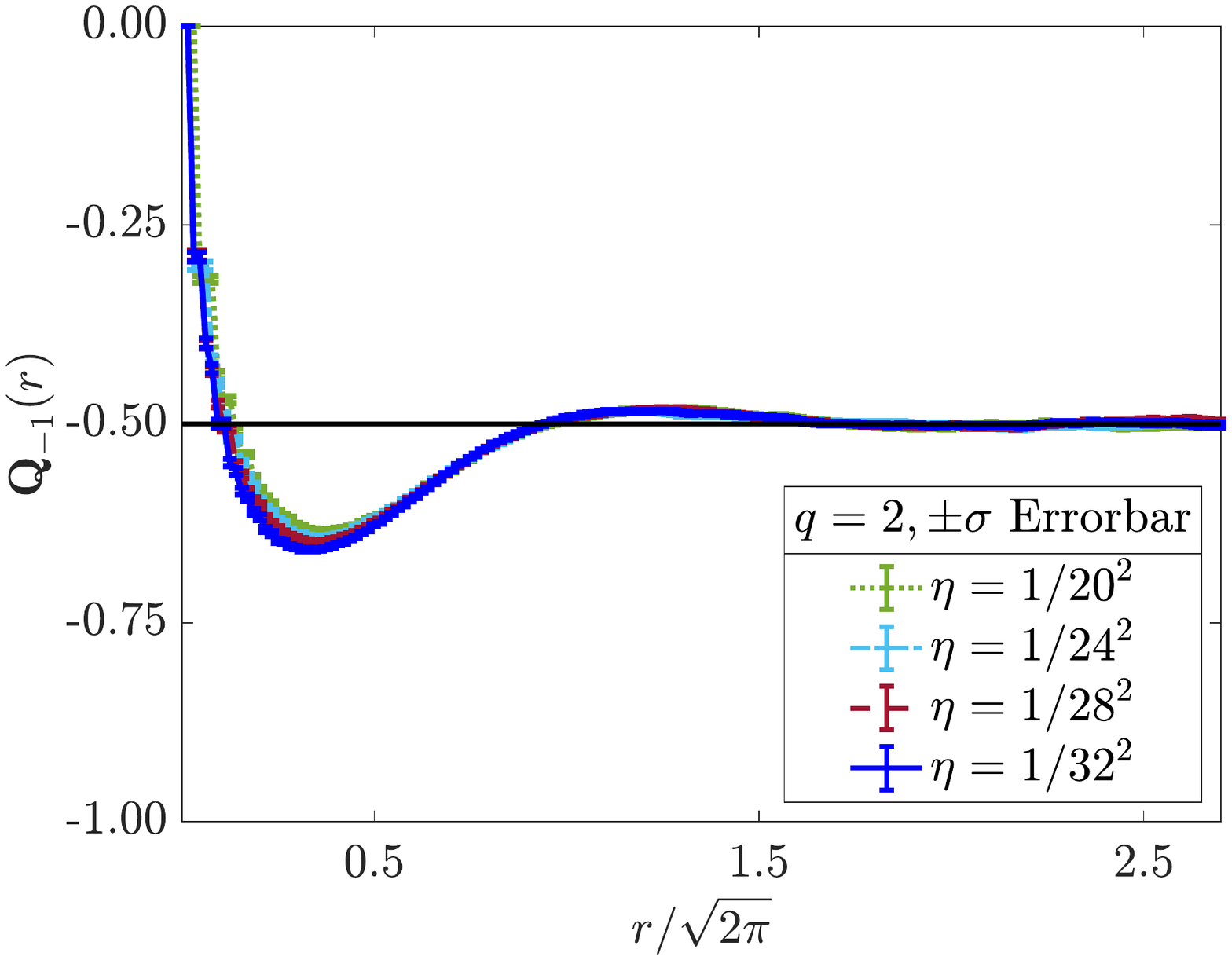}\quad
\raisebox{5cm}{\bf(d)} \includegraphics[trim={0.7cm 7cm 0.7cm 7cm},clip, scale=0.38]{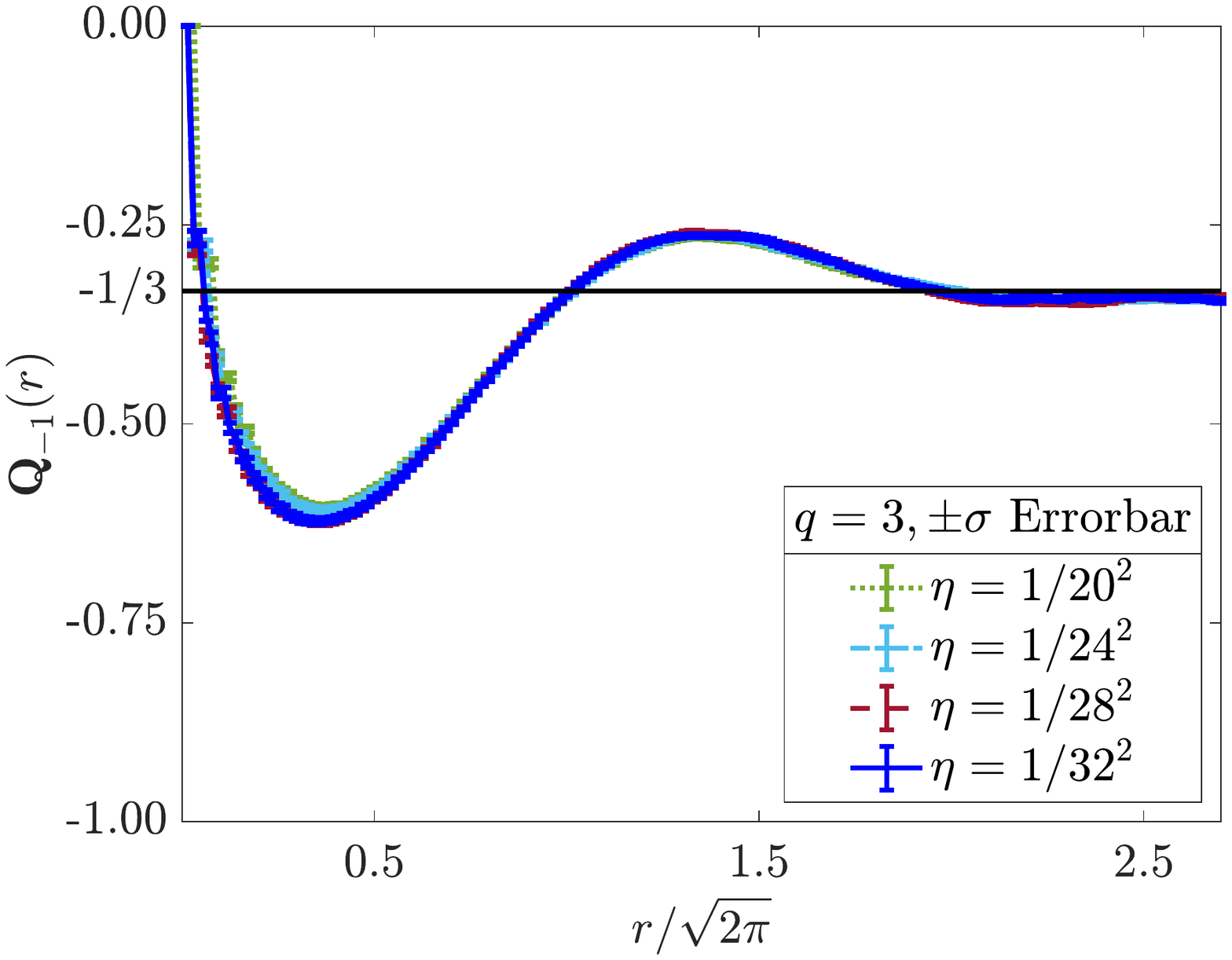}
\caption{Excess charge $\textbf{Q}_{-1}(r)$ of the quasielectron \re{Latt_1QE} for inverse filling fractions $q = 2$ [\textbf{(a)} and \textbf{(c)}] and $q = 3$ [\textbf{(b)} and \textbf{(d)}]. The quasielectron is at the origin of a circular disk of radius $R \approx 5.6\sqrt{2\pi}$ with a square lattice with $N$ lattice points embedded on it. We define $\textbf{Q}_{-1}(r)$ as a function of the radial distance $r$ from the quasielectron in the units of absolute electronic charge $e$ in \Eref{Exc_Charge}. We show the $\textbf{Q}_{-1}(r)$ averaged over $25$ different Monte Carlo datasets in \textbf{(a)} and \textbf{(b)}. Additionally, in \textbf{(c)} and \textbf{(d)}, we include an errorbar of $\pm \sigma$ for the excess charge profiles, where $\sigma$ is the variance of $\textbf{Q}_{-1}(r)$. To approach the continuum, starting from the lattice limit $\eta = 1$, in consecutive steps we decrease $\eta$ as $\eta = 1/L^{2}$, and put $L^{2}N$ lattice sites on the same circular disk of radius $R$. We show the $\textbf{Q}_{-1}(r)$ profiles for $L = 20, 24, 28,$ and $32$. For $r\gtrsim 2\sqrt{2\pi}$, for all the $\eta$ values shown here, the excess charge for the quasielectron at filling fractions $1/q$ is indeed close to $1/q$. When $r$ is small, i.e., near the trough of the excess charge profile, $\textbf{Q}_{-1}(r)$ goes down as we decrease $\eta$. It is hard to conclude if the profile has reached a limit or not after including the $\pm \sigma$ errorbar. In \Sref{Sec: Cont_QE_Analytical_Argument}, we argue that the curves do not converge.}
\label{Square_Scaling}
\end{indented} 
\end{figure}

\begin{figure}[tbp!]
\begin{indented}\item[]
\fl \raisebox{5cm}{\bf(a)} \includegraphics[trim={0.7cm 7cm 0.7cm 7cm},clip, scale=0.38]{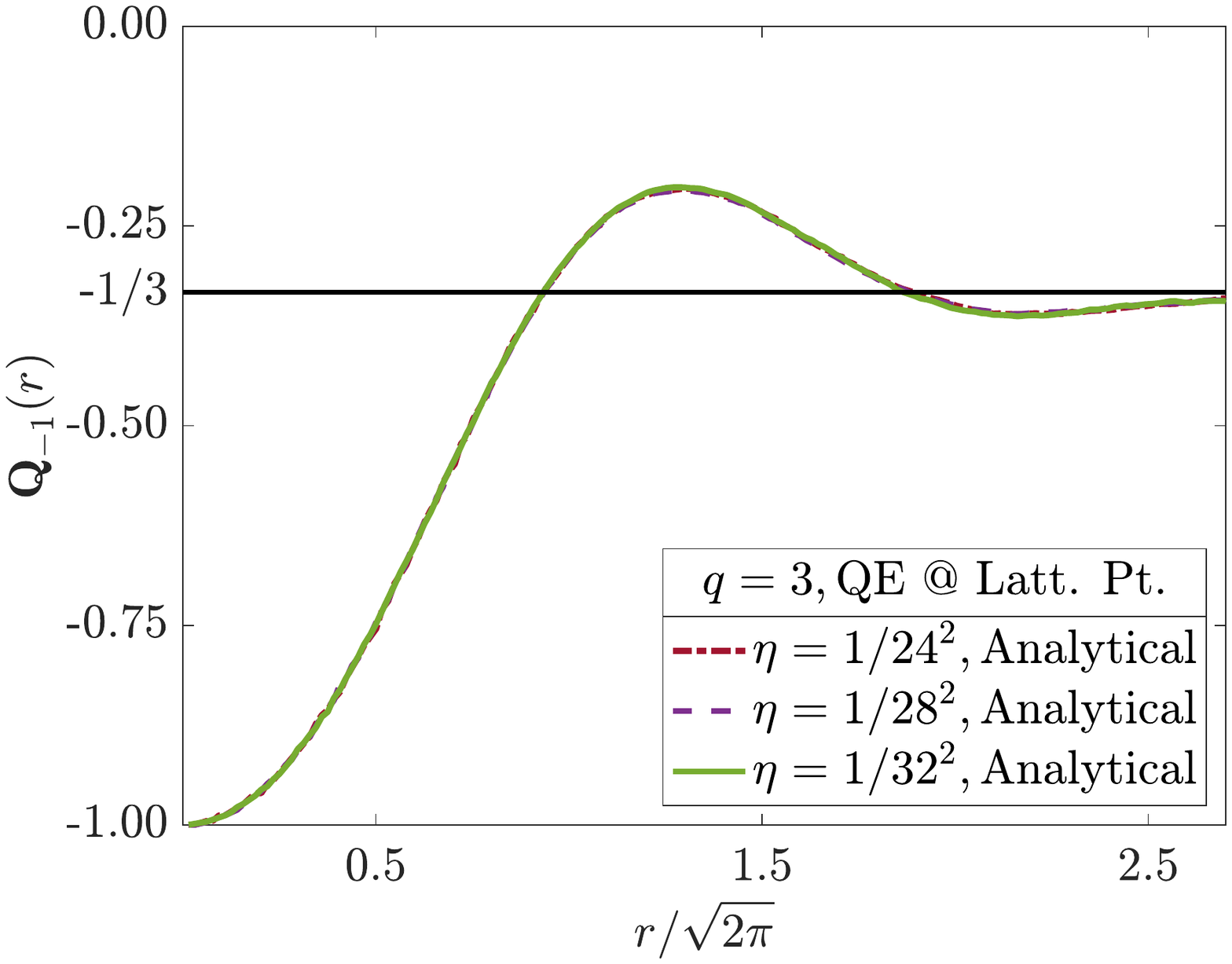}\quad
\raisebox{5cm}{\bf(b)} \includegraphics[trim={0.7cm 7cm 0.7cm 7cm},clip, scale=0.38]{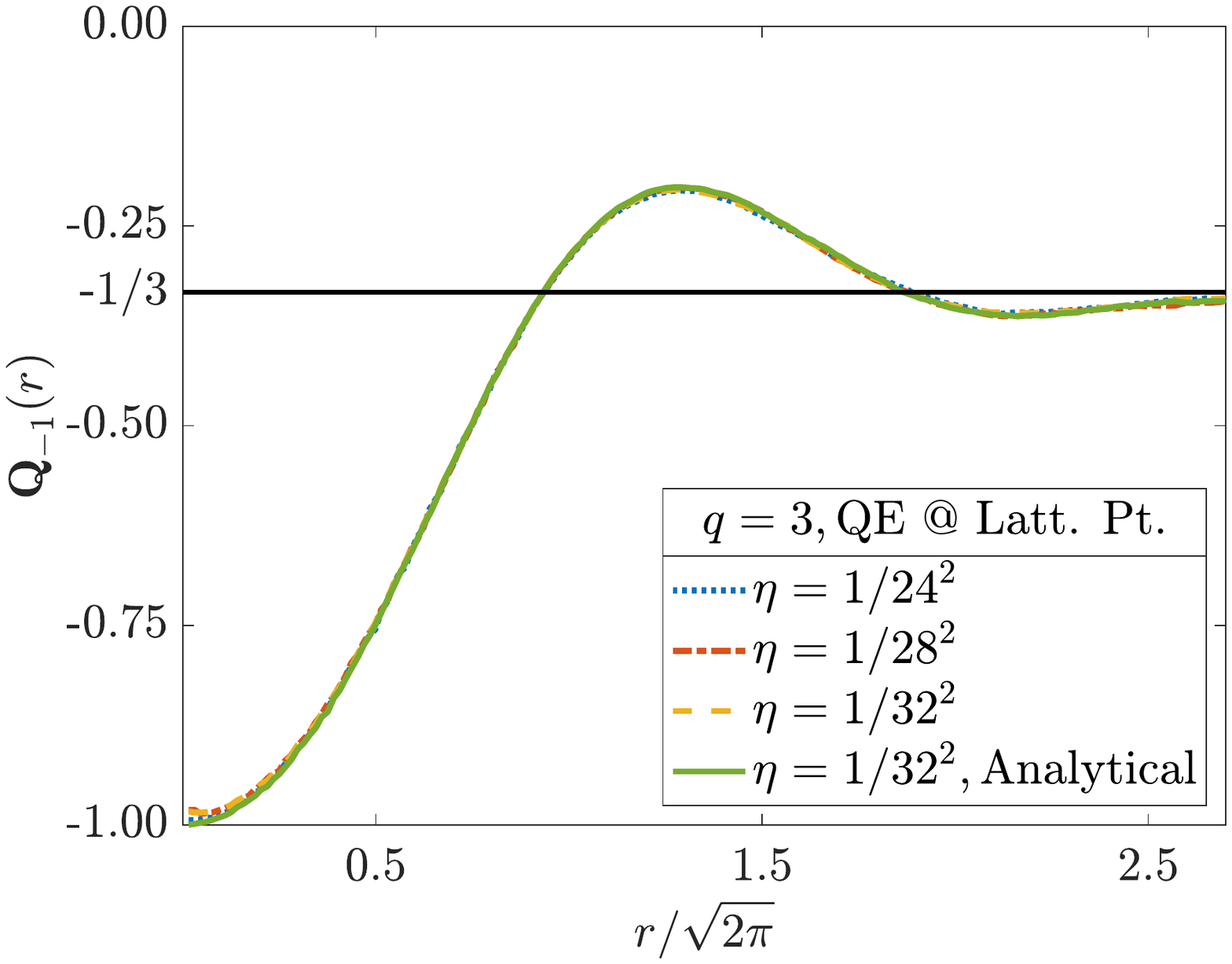} \\
\fl \raisebox{5cm}{\bf(c)} \includegraphics[trim={0.7cm 7cm 0.7cm 7cm},clip, scale=0.38]{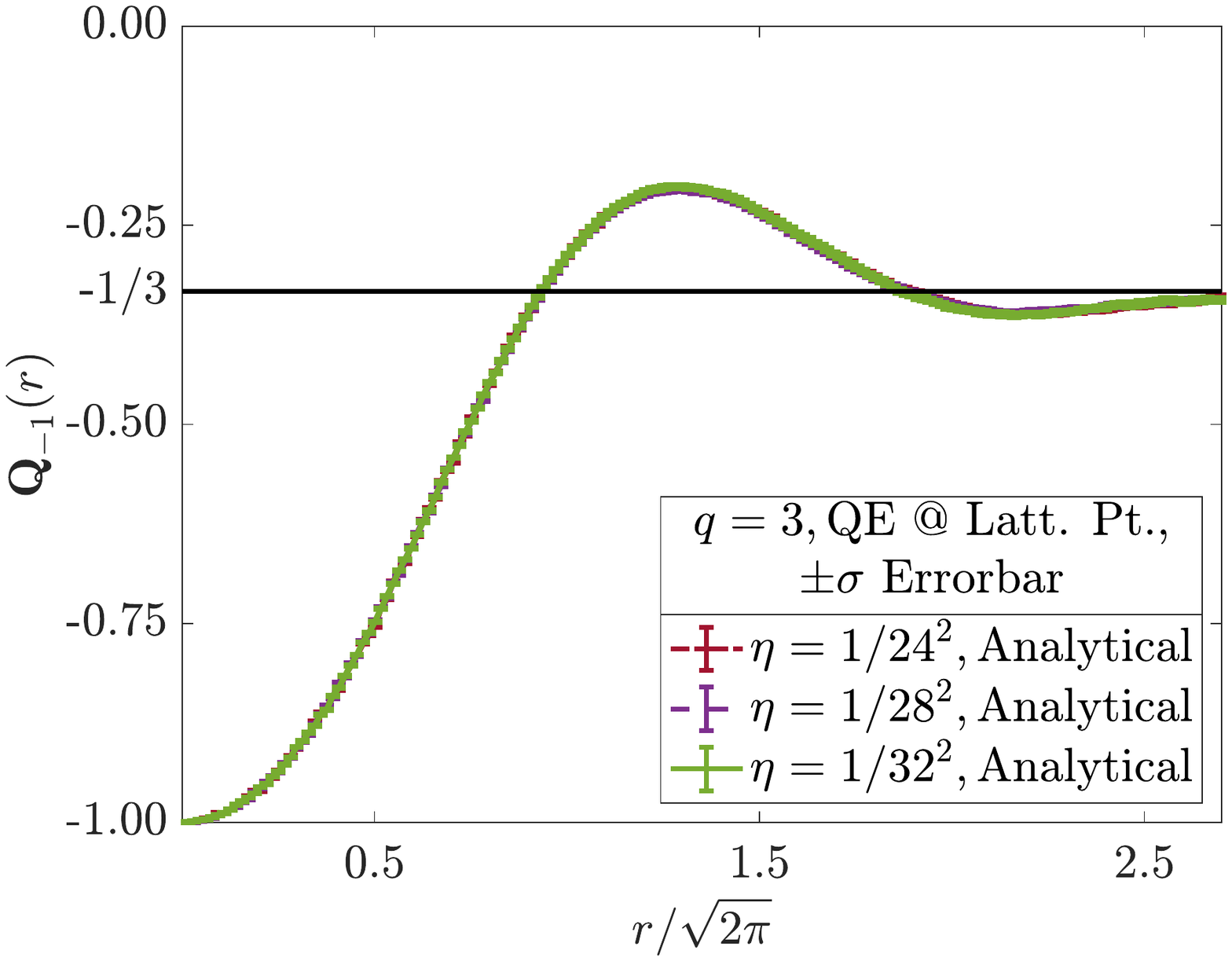}\quad
\raisebox{5cm}{\bf(d)} \includegraphics[trim={0.7cm 7cm 0.7cm 7cm},clip, scale=0.38]{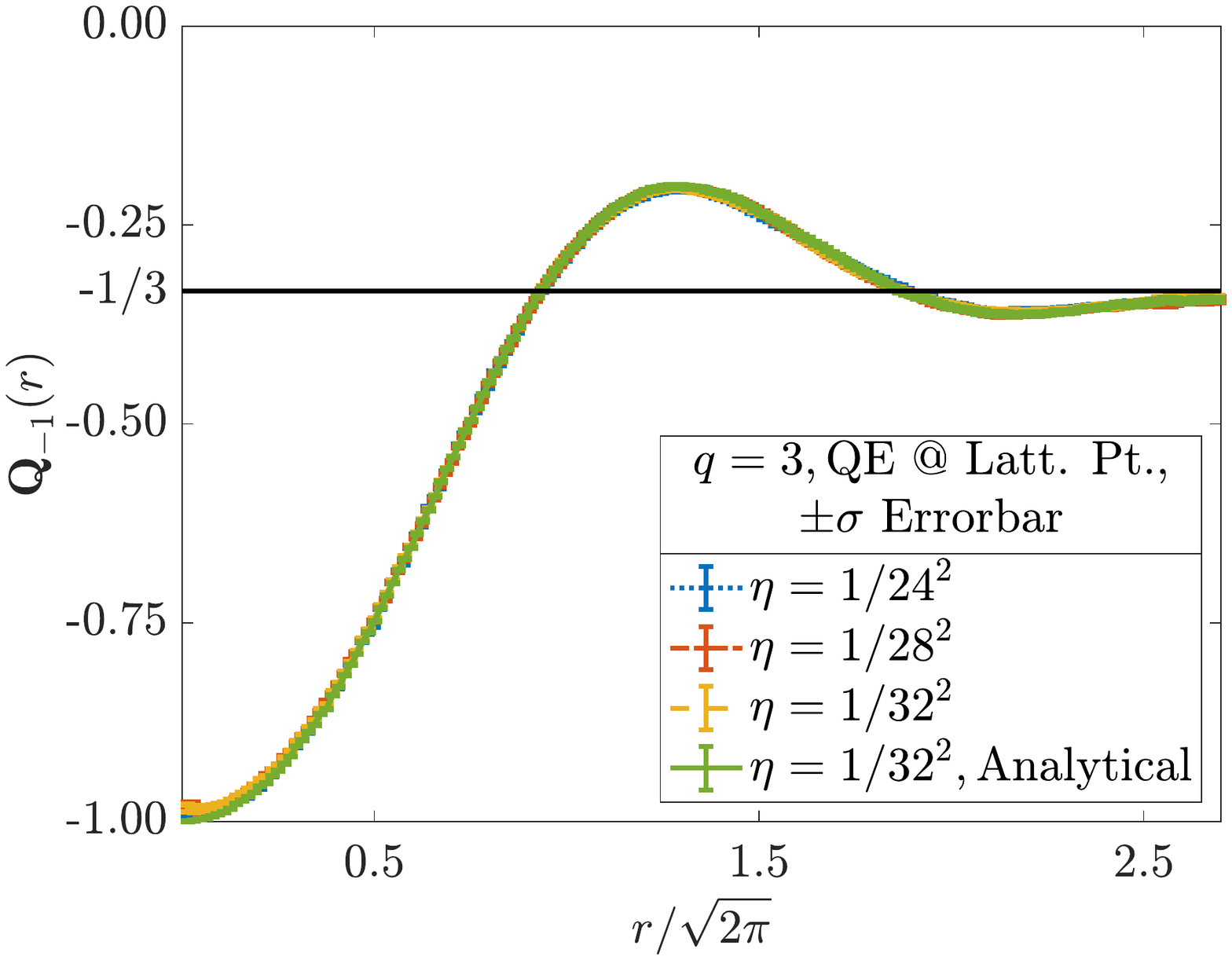}
\caption{Excess charge $\textbf{Q}_{-1}(r)$ profile with the quasielectron on a lattice site for $q = 3$. In \textbf{(a)}, we plot  $\textbf{Q}_{-1}(r)$ for $\eta = 1/24^{2}, 1/28^{2},$ and $1/32^{2}$ using the exact analytical wavefunction, where we always keep the lattice site at $\sqrt{2\pi\eta}\left(1/2, 1/2\right)$ occupied. The convergence of $\textbf{Q}_{-1}(r)$ is clearly demonstrated here. In \textbf{(b)}, we compare the excess charge profile for $\eta = 1/32^{2}$ obtained from the analytical wavefunction to the ones, with $\eta = 1/24^{2}, 1/28^{2},$ and $1/32^{2}$, obtained from the wavefunction with the quasielectron close to a lattice site at $w = \sqrt{2\pi}\left(\sqrt{\eta}/2, \sqrt{\eta}/2 -10^{-4}\right)$. In \textbf{(c)} and \textbf{(d)}, we show \textbf{(a)} and \textbf{(b)} with $\pm\sigma$ errorbars, respectively.}
\label{Near_Latt_Pt}
\end{indented} 
\end{figure}

\begin{figure}[tbp!]
\begin{indented}\item[]
\includegraphics[trim={0.7cm 7cm 0.7cm 7cm},clip, scale=0.38]{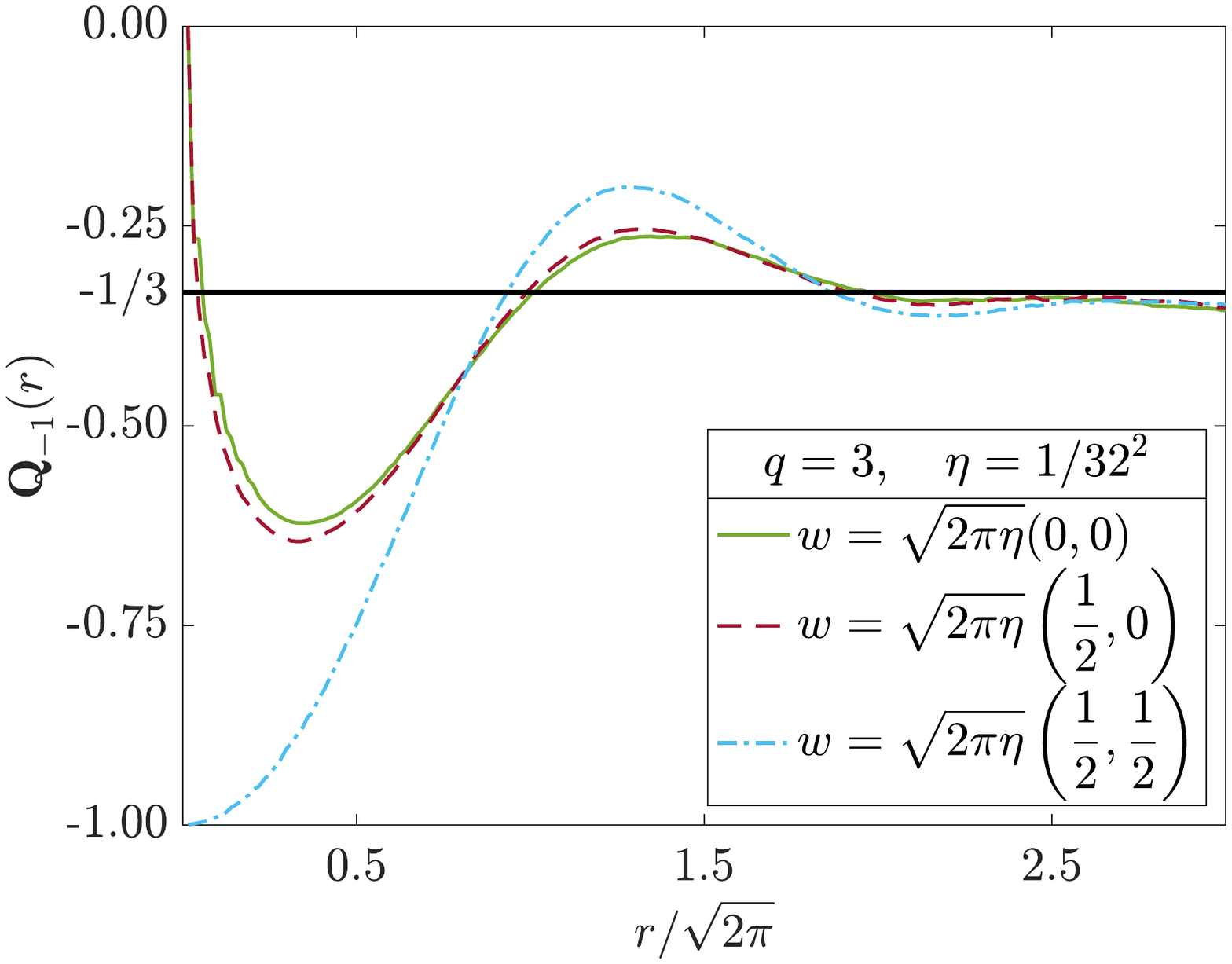}
\caption{Excess charge profiles for $q = 3$ and $\eta = 1/32^{2}$ with three different inequivalent positions $w$ of the quasielectron in the square lattice. Here $r$ is the radial distance from the quasielectron position. The three positions -- center of a square plaquette (solid green), midpoint of a side (brown dashed), and on top of a lattice site (indigo dash-dot) -- lead to different $\textbf{Q}_{-1}(r)$ profiles.}
\label{Diff_Pos}
\end{indented}
\end{figure}


We then proceed to study the continuum limit of the quasielectron ansatz \re{Latt_1QE} for a square lattice defined on a circular disk of radius $R \approx 5.6\sqrt{2\pi}$. Without loss of generality, we consider the following initial primitive lattice vectors:
\beg
\bm{a}_{1} = \hat{x}\sqrt{2\pi}, \qquad \bm{a}_{2} = \hat{y}\sqrt{2\pi},
\label{INIT_Sq_Latt}
\en 
where $\hat{x}$ and $\hat{y}$ are the unit vectors in the $x$ and $y$ directions. Since the area per primitive cell needs to be $2\pi\eta$, we initially have $\eta = 1$. The circular disk initially contains $96$ lattice points.

We gradually decrease $\eta$ while keeping $M$ and $N\eta$ fixed. To achieve this, in the next step we consider a square lattice in which we replace each of the unit cells in the original lattice with $L^{2}$ smaller unit cells each having the area $2\pi/L^{2}$. In this way one has
\beg
N = N_{\mathrm{New}}/L^{2}, \quad \eta = L^{2}\eta_{\mathrm{New}}, \quad  N_{\mathrm{New}}\eta_{\mathrm{New}} = N\eta.
\en

As shown in \Fref{Lattice_Doubling}, we shall consider three distinct positions of the quasielectron -- \textbf{(1)} the center of a square plaquette formed by four lattice sites, \textbf{(2)} midway between two lattice sites, and \textbf{(3)} on a lattice point. We place the origin at \textbf{(1)}. As we decrease $\eta$ and increase $N$ while keeping $N\eta$ constant, \textbf{(2)} and \textbf{(3)} converge to the origin. In the continuum limit, all the three inequivalent quasielectron positions coincide.

We already explained below \Eref{Norm_Sq} that putting the quasielectron on top of a lattice site [position \textbf{(3)}] obtains a finite trial wavefunction. That explanation also holds in the continuum limit. However, such reasoning does not work, when the quasielectron is at the center of a square plaquette [position \textbf{(1)}], or midway between two lattice sites [position \textbf{(2)}]. The question is if the continuum trial wavefunctions, with the quasielectron in the above three inequivalent positions, coincide and converge. At the continuum limit, one defines the LLL. Laughlin states and their quasiparticle excitations should be confined to it. We need to verify this as well.

We study the excess charge profile, which is defined as follows:
\beg
\textbf{Q}_{-1} (r) = \sum_{i = 1}^{N} \Theta\left(r - |z_{i}| \right)\left(\mean{n_{i}}_{0} - \mean{n_{i}}_{-1}\right).
\label{Exc_Charge}
\en
Here $\Theta(x)$ denotes the Heaviside step function
\begin{eqnarray}
\Theta(x) = \left\{\begin{array}{cl}0 & \textrm{for }x<0\\
1 & \textrm{otherwise}\end{array}\right..
\label{Step_Fn}
\end{eqnarray}
In \Eref{Exc_Charge}, $\mean{n_{i}}_{-1}$ are the occupancies of the lattice sites with the quasielectron ansatz \re{Latt_1QE}, and $\mean{n_{i}}_{0}$ are the occupancies of the lattice sites calculated with the wavefunction \re{Latt_QP} and all $p_{k} = 0$. In \Fsref{Loc_QE}\blue{a} and \ref{Loc_QE}\blue{b}, we show the excess particle densities $\left(\mean{n_{i}}_{-1} - \mean{n_{i}}_{0}\right)$ for $q = 2$ and $q = 3$.

We have considered a quasielectron at the center of a square plaquette down to $\eta = 1/32^{2}$ for both $q = 2$ and~$3$ in \Fref{Square_Scaling}. Here we go to much smaller values of $\eta$ than what was considered in Ref.~\cite{anqe}. The smallest value considered there was $\eta = 1/10^2$. The excess charge, as expected, saturates to $\textbf{Q}_{-1} (r) = -1/q$ in both \Fsref{Square_Scaling}\blue{a} and~\ref{Square_Scaling}\blue{b} for $r \gtrsim 2.0\sqrt{2\pi}$. For distances smaller than $\sqrt{2\pi}$, especially near the trough of the excess charge profile, it keeps going down as we decrease $\eta$. However, the amount of change is so small -- keeping in mind the numerical accuracy of our calculations (see \Fsref{Square_Scaling}\blue{c} and~\ref{Square_Scaling}\blue{d}, where we have included the $\pm \sigma$ errorbar) -- that it is difficult to comment whether the charge profile has saturated or not.

On the other hand in \Fsref{Near_Latt_Pt}\blue{a} and \ref{Near_Latt_Pt}\blue{c} (in the latter we have included $\pm \sigma$ errorbars), we show the converged (unlike \Fref{Square_Scaling}) excess charge profiles for small values of $\eta$ using the exact ansatz wavefunction for the quasielectron position $w =\sqrt{2\pi\eta}\left(1/2, 1/2\right)$. We keep the above lattice site always occupied to obtain $\textbf{Q}_{-1} (r)$ with the exact ansatz while performing the Monte Carlo simulations. In \Fsref{Near_Latt_Pt}\blue{b} and \ref{Near_Latt_Pt}\blue{d} (the latter shows $\pm \sigma$ errorbars), we compare $\textbf{Q}_{-1} (r)$ calculated with the exact analytical ansatz to the the one obtained by keeping the quasielectron close to a lattice point $\sqrt{2\pi\eta}\left(1/2, 1/2\right)$. In \Fref{Diff_Pos}, we plot the $\textbf{Q}_{-1} (r)$ for the three different quasielectron positions shown in \Fref{Lattice_Doubling}.

\Fsref{Loc_QE}, \ref{Square_Scaling}, \ref{Near_Latt_Pt}, and \ref{Diff_Pos} demonstrate that the quasielectron is indeed screened. Combining this with the arguments of \Sref{Sec: Braiding_Lattice_Wavefunc} shows that the quasielectron described by \Eref{Latt_QP} also has the braiding properties expected for anyonic quasielectrons.

\subsection{Divergence of the Continuum Quasielectron: Analytical Argument}
\label{Sec: Cont_QE_Analytical_Argument}


\begin{figure}[tbp!]
\begin{indented}\item[]
\fl \raisebox{5cm}{\bf(a)} \includegraphics[trim={0.7cm 7cm 0.7cm 7cm},clip, scale=0.38]{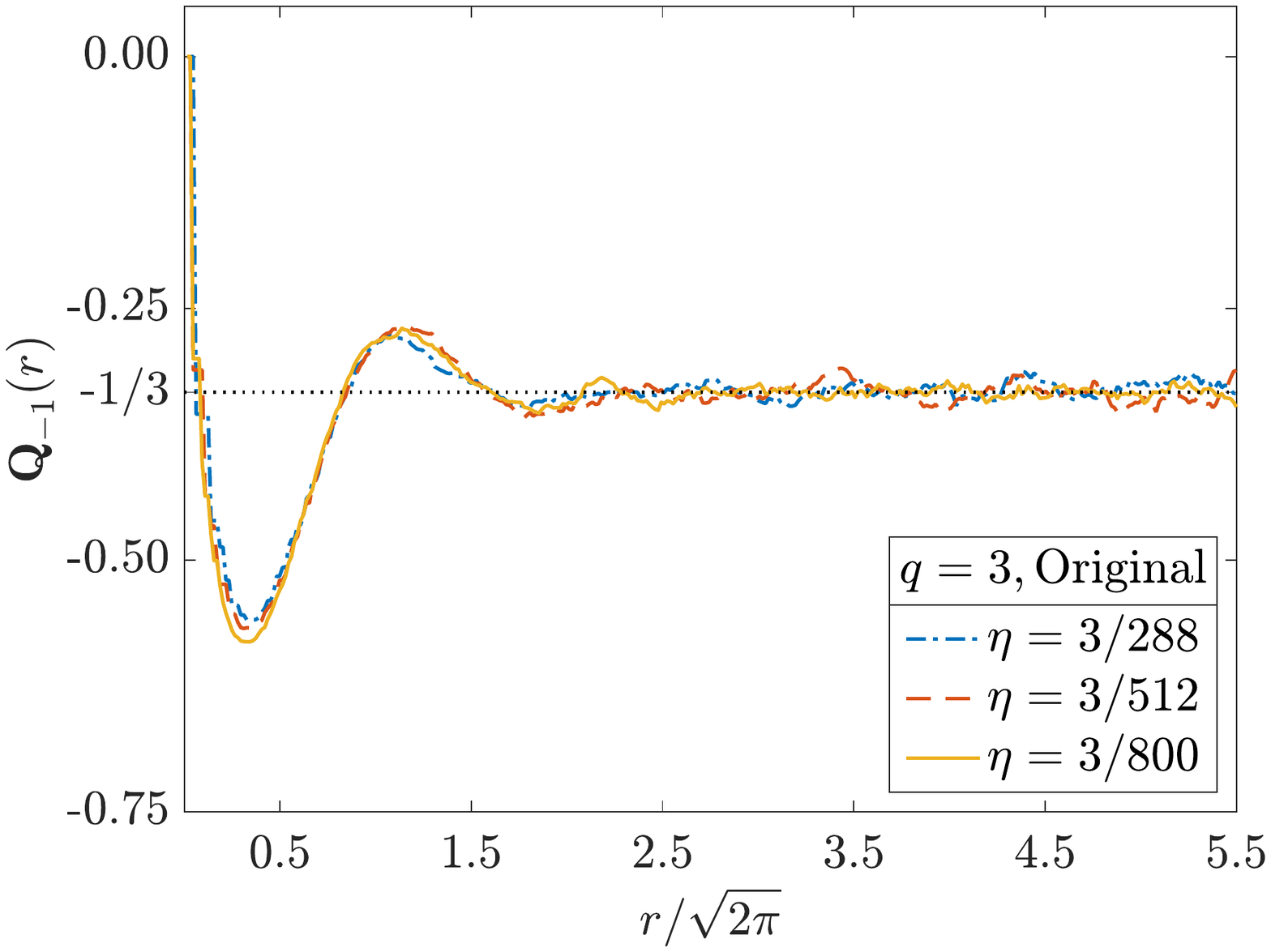}\llap{\raisebox{3.5cm}{\includegraphics[trim={0.5cm 7cm 1cm 7cm}, clip, height=2.6cm]{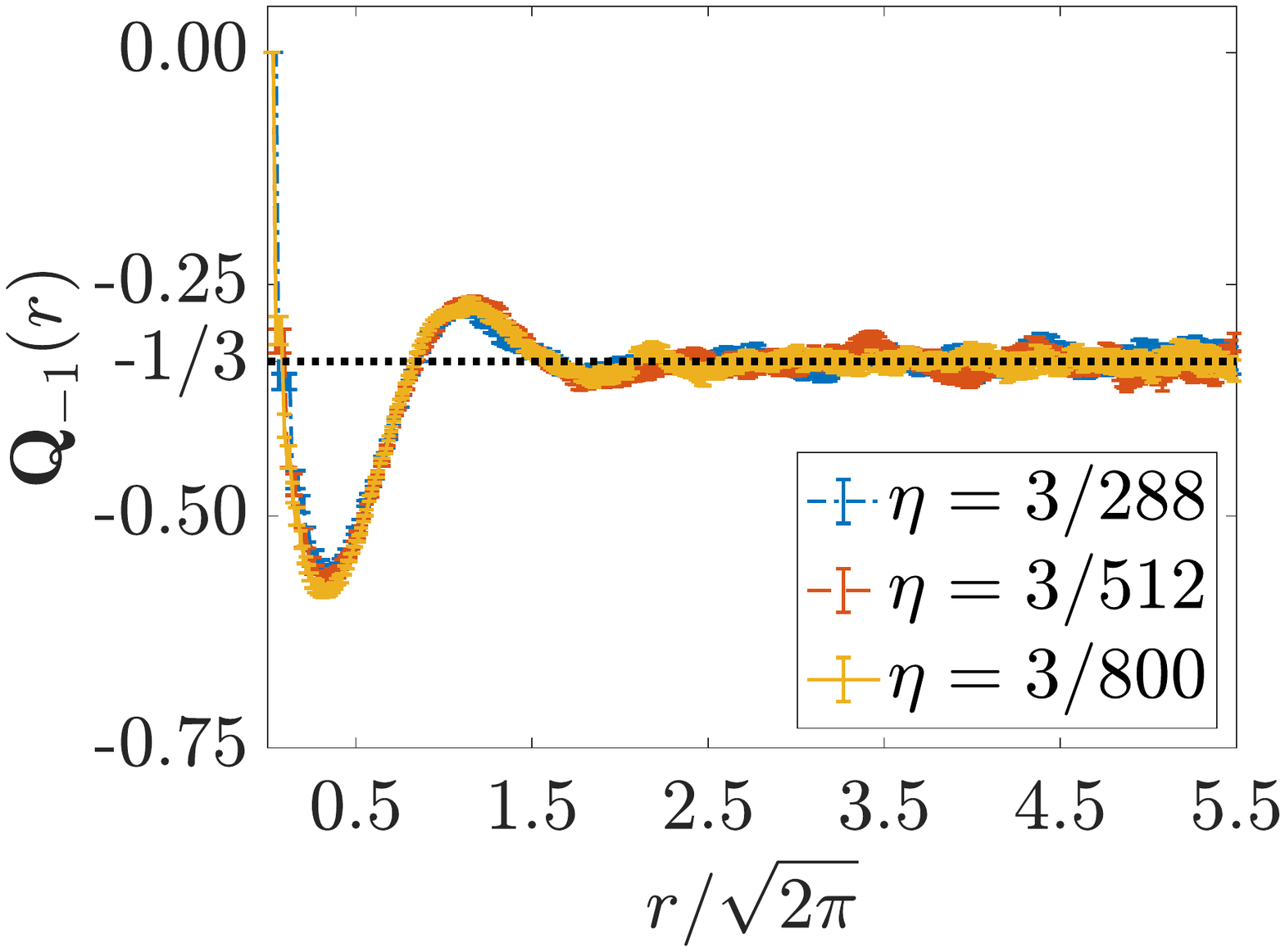}}}\quad
\raisebox{5cm}{\bf(b)} \includegraphics[trim={0.7cm 7cm 0.7cm 7cm},clip, scale=0.38] {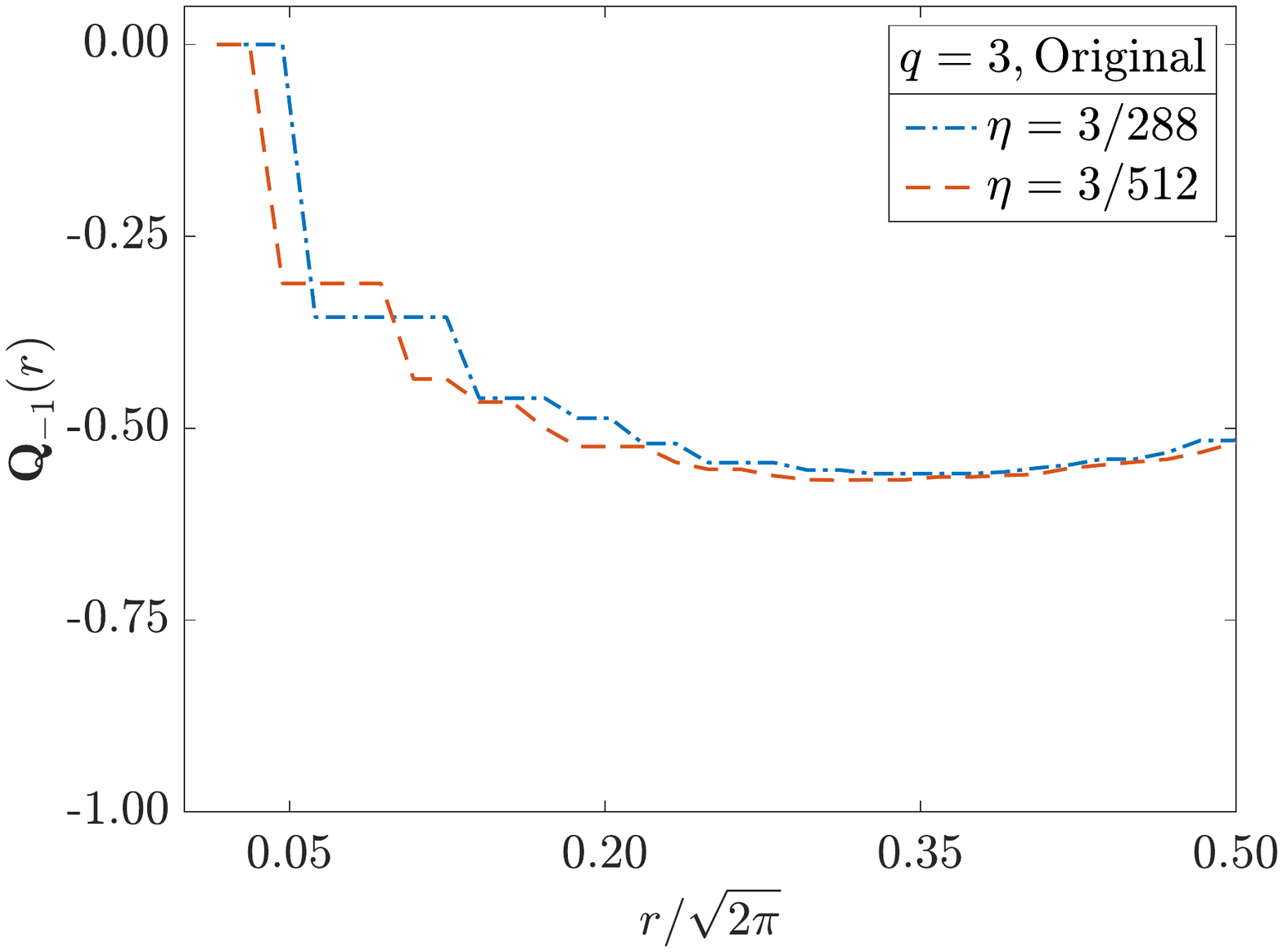}\\
\fl \raisebox{5cm}{\bf(c)} \includegraphics[trim={0.7cm 7cm 0.7cm 7cm},clip, scale=0.38]{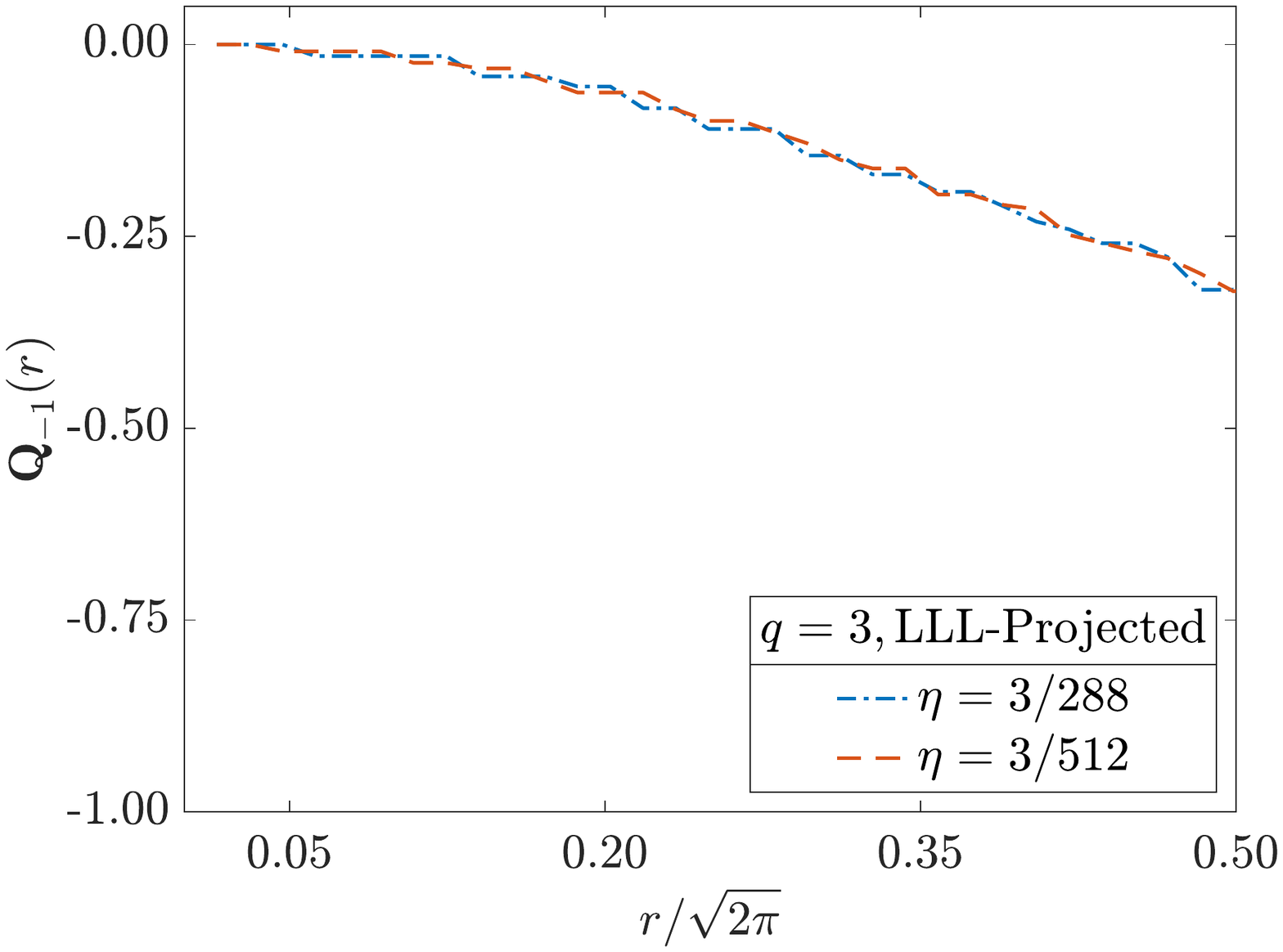}
\begin{picture}(0,0)
\put(-180,59){\includegraphics[trim={0.7cm 7cm 0.7cm 7cm},clip,height=2.6cm]{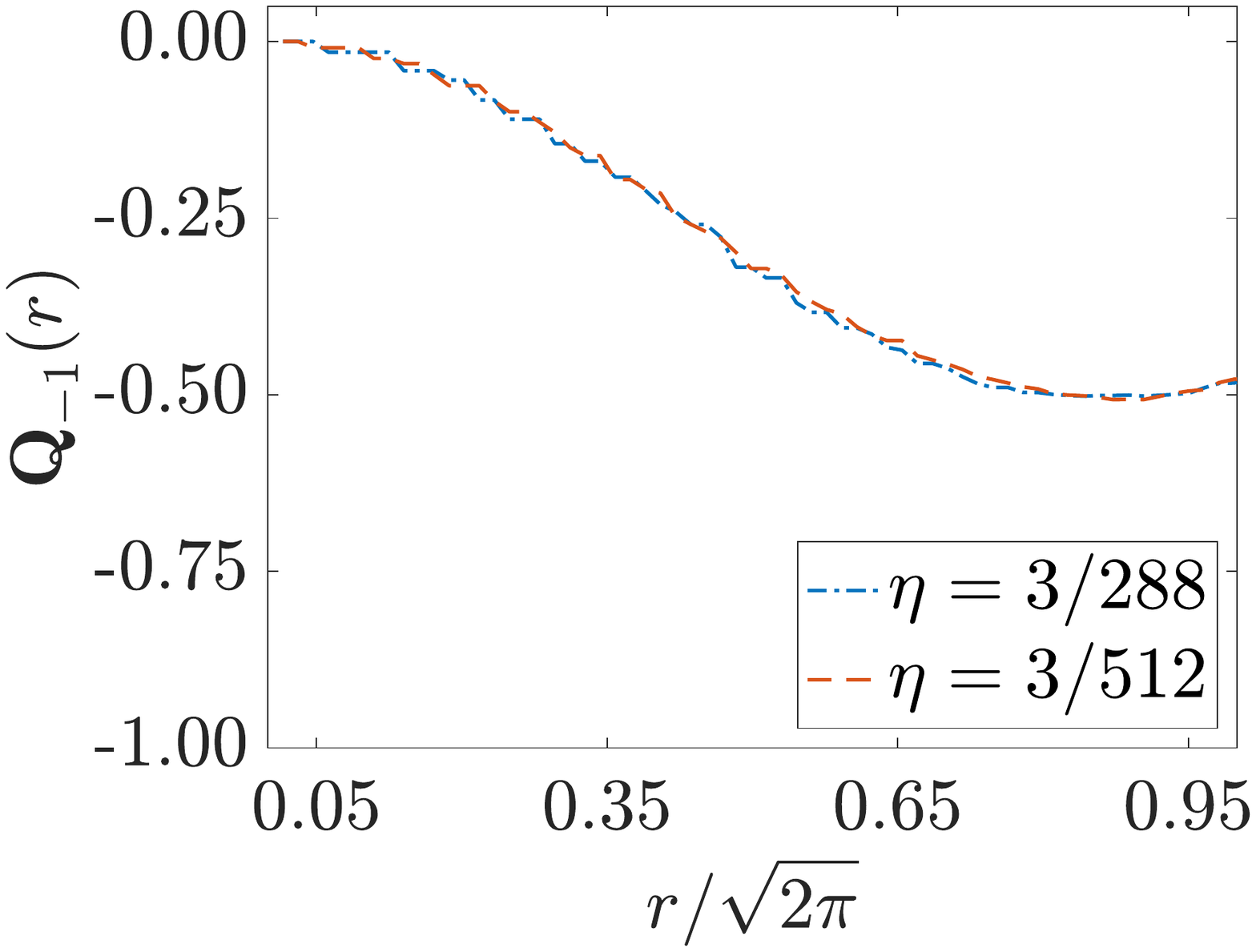}}
\end{picture}
\quad
\raisebox{5cm}{\bf(c)} \includegraphics[trim={0.7cm 7cm 0.7cm 7cm},clip, scale=0.38]{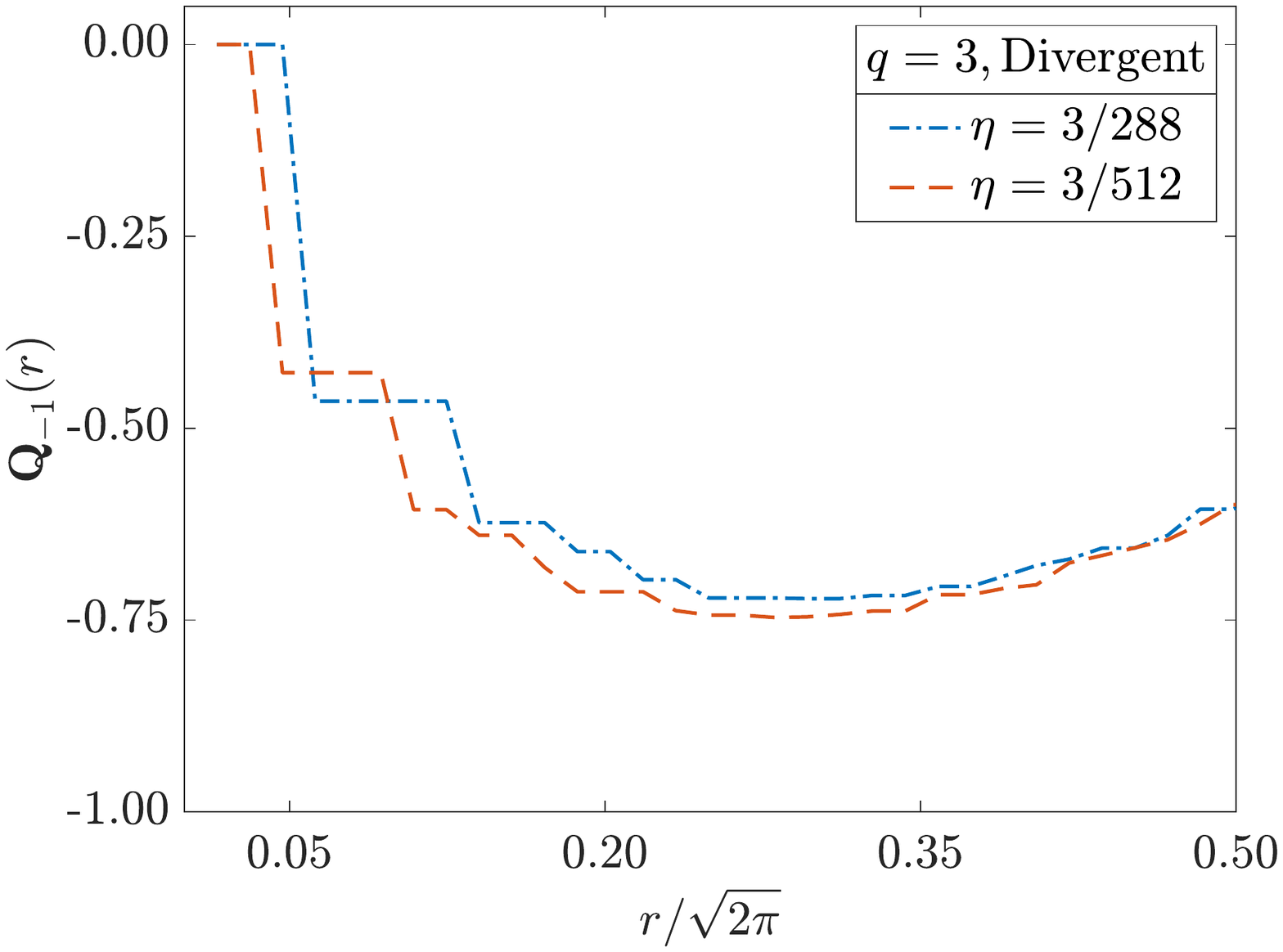}
\caption{The divergent terms in the expansion of \Eref{Latt_1QE} [the terms that are not confined to the LLL in the continuum -- i.e., the ones left after subtracting \Eref{Latt_1QE_Mod} from \re{Latt_1QE}] are responsible for the changes in $\textbf{Q}_{-1}(r)$ between $0<r\lesssim \sqrt{2\pi}$. All of these plots are for a square lattice on a circular disk of radius $R \approx 8.3\sqrt{2\pi}$ and $q = 3$. We plot $\textbf{Q}_{-1}(r)$ vs.\ $r$ with ansatz \re{Latt_1QE} for $\eta = 3/288, 3/512,$ and $3/800$ in \textbf{(a)}. As we decrease $\eta$, the minimum  of $\textbf{Q}_{-1}(r)$ near $r = 0.25\sqrt{2\pi}$ keeps decreasing. The oscillations about $\textbf{Q}_{-1}(r) = -1/3$ for $r \gtrsim 2.0\sqrt{2\pi}$ gradually decrease with decreasing $\eta$. We have included the $\pm\sigma$ errorbars in the inset of \textbf{(a)}. In \textbf{(b)}, \textbf{(c)}, and \textbf{(d)} we show the $\textbf{Q}_{-1}(r)$ vs.\ $r$ plots for $\eta = 3/288$ and $3/512$, where we compute the excess charges with the ansatz \re{Latt_1QE}, with ansatz \re{Latt_1QE} projected to the LLL [i.e., with ansatz \re{Latt_1QE_Mod}], and with only the divergent terms in \Eref{Latt_1QE}, respectively. From \textbf{(c)}, we notice that $\textbf{Q}_{-1}(r)$ calculated with \Eref{Latt_1QE_Mod} for the two different values of $\eta$ are almost identical. The inset of \textbf{(c)} shows $\textbf{Q}_{-1}(r)$ for upto $r = \sqrt{2\pi}$. In both \textbf{(b)} and \textbf{(d)}, there are significant changes in $\textbf{Q}_{-1}(r)$ when $\eta$ changes.}
\label{Div_Terms_Responsible}
\end{indented} 
\end{figure}


The foregoing numerical calculation could not definitively answer if the quasielectron \re{Latt_1QE} leads to a singularity in the continuum, when it is placed at the center of a square plaquette or midway between two lattice points. We therefore take recourse to an analytical argument. When we are close to the continuum limit, there are so many lattice points that the electrons can be almost anywhere. However, at the place where we have the singularity, the difference between the lattice and the continuum is important.

In particular, the underlying lattice structure introduces a distance of closest approach from the quasielectron position $\xi$ (as long as the quasielectron is not placed on top of a lattice point).  Since the length of the primitive lattice vectors in the square lattice is $\sqrt{2\pi\eta}$, we have
\beg
\xi \propto \sqrt{\eta} \propto 1/\sqrt{N}.
\label{Dist_Closes_Appr}
\en
The proportionality constant in \Eref{Dist_Closes_Appr} depends on the position of the quasielectron with respect to a square plaquette.

In the continuum, the LLL is well-defined and the quasielectron should be entirely confined to it. This does not hold for the wavefunction \re{Latt_1QE} because of the $1/z_{i}$ factors. We write the norm of the single electron LLL wavefunction $\rme^{-|Z|^{2}/4}$ multiplied by the quasielectron pole $1/Z$ over a domain with a hole around the singularity as follows:
\beg 
\int\int_{D^{\prime}} \bigg|\frac{\rme^{-|Z|^{2}/4}}{Z}\bigg|^{2} d^{2}Z = \int\int_{D} \Theta^{2}\left(|Z| - \xi\right) \bigg|\frac{\rme^{-|Z|^{2}/4}}{Z}\bigg|^{2} d^{2}Z,
\label{HLL_Int}
\en 
where $\Theta(x)$ is the Heaviside step function \re{Step_Fn}. We perform the integral in the left hand side on a sufficiently large disk $D$ with a small hole of radius $\xi$ around the position of the quasielectron -- the origin. We denote this not-simply-connected domain as $D^{\prime}$. The integral on the right hand side, with the wavefunction $\Theta\left(|Z| - \xi\right)\rme^{-|Z|^{2}/4}/Z$, is performed on the full disk. The wavefunctions in both sides of \Eref{HLL_Int} do not live in the LLL.

The preceding discussion establishes that the lattice quasielectron \re{Latt_1QE} tends to the following continuum quasielectron living on $D^{\prime}$:
\beg 
\fl \big|\psi_{\mathrm{QE}}\big\rangle_{\mathrm{cont}}  = \nc^{-1}\int\cdots\int_{D^{\prime^{M}}} \prod_{i = 1}^{M}\Diff2 Z_{i} \big|\Zv\big\rangle \prod_{i} Z_{i}^{-1}\prod_{i<j} \left(Z_{i} - Z_{j}\right)^{q} \rme^{-\sum_{i = 1}^{M}|Z_{i}|^2/4l_{B}^2}.
\label{Laugh_1QE}
\en 
Expanding the product $\prod_{i<j}\left(Z_{i} - Z_{j}\right)^{q}$, we write the wavefunction \re{Laugh_1QE} as follows:
\beg 
\fl \big|\psi_{\mathrm{QE}}\big\rangle_{\mathrm{cont}} = \nc^{-1}\int\cdots\int_{D^{\prime^{M}}} \prod_{i = 1}^{M}\Diff2 Z_{i} \big|\Zv\big\rangle \sum_{\qv} \mathcal{E}_{\qv} \prod_{l = 1}^{M}Z_{l}^{q_{l}-1}\rme^{-\frac{|Z_{l}|^2}{4}},
\label{Exp_Inv_QH}
\en 
where $\mathcal{E}_{\qv}$ are different expansion coefficients, and $\qv$ satisfy
\beg
\sum_{l}q_{l} = qM(M-1)/2,\quad 0 \leqslant q_{l}  \leqslant (M-1)q.
\label{mk_constr}
\en

Note that $\mathcal{E}_{\qv}$ are the same coefficients that appear in the expansion of the Laughlin wavefunction (\Eref{Laugh_QP} with all $p_{k} = 0$) in the basis $\big|\qv\big\rangle$ of single electron wavefunction product, which we write as follows:
\beg
\big |\psi\big\rangle = \nc^{-1} \sum_{\qv} \mathcal{E}_{\qv} \big|\qv\big\rangle.
\label{Laugh_WF_SEB}
\en
To go back to the coordinate basis $\big|\Zv\big\rangle$, one needs to insert the following identity operator,
\beg
\mathds{1} = \int\cdots\int_{D^{M}} \prod_{i = 1}^{M}\Diff2 Z_{i} \big|\Zv\big\rangle \big\langle \Zv \big|,
\en
in \Eref{Laugh_WF_SEB}, and use
\beg
\big\langle \Zv \big| \qv\big\rangle \propto \prod_{l = 1}^{M}Z_{l}^{q_{l}}\rme^{-\frac{|Z_{l}|^2}{4}}.
\en
Using this, we write \Eref{Exp_Inv_QH} as:
\beg
\big|\psi_{\mathrm{QE}}\big\rangle_{\mathrm{cont}} = \nc^{-1} \sum_{\qv} \mathcal{E}_{\qv} \big|q_{1} - 1, \ldots q_{M} - 1\big\rangle,
\label{Exp_Inv_QH_SEB}
\en
where $\big|q_{1} - 1, \ldots q_{M} - 1\big\rangle$ is an orthonormal product basis similar to $\big|\qv\big\rangle$. This new single electron basis on $D^{\prime}$, however, includes the element $\big|-1\big\rangle$ with
\beg
\big\langle Z \big| -1 \big\rangle \propto Z^{-1}\rme^{-\frac{|Z|^2}{4}},
\label{Pole_Single_Electron}
\en
to accommodate the quasielectron pole. Note that the function \re{Pole_Single_Electron} is normalizable on the domain ${D^{\prime}}$.

We have two types of terms in the expansion \re{Exp_Inv_QH} -- the ones with all the $q_{l}$ positive, and the ones with at least one of them equal to zero. The first type of terms $\lbrace\rqv\rbrace$ do not lead to any singularity. We delineate this regular (not normalized) part of the continuum quasielectron wavefunction \re{Exp_Inv_QH} as follows:
\beg
\fl \big |\psi^{\mathrm{reg}}_{\mathrm{QE}}\big\rangle_{\mathrm{cont}} = \nc^{-1}\int\cdots\int_{D^{\prime^{M}}} \prod_{i = 1}^{M}\Diff2 Z_{i} \big|\Zv\big\rangle \sum_{\rqv} \mathcal{E}_{\rqv}\prod_{l = 1}^{M}Z_{l}^{q^{\mathrm{reg}}_{l}-1}\rme^{-\frac{|Z_{l}|^2}{4}},
\label{Exp_Inv_QH_Reg}
\en 
where $\nc$ is the same as in \Esref{Laugh_1QE} and \re{Exp_Inv_QH}.

The second type of terms $\lbrace\sqv\rbrace$ however, have at least one pole at the origin. Among all these terms, those that have poles of the type $Z_{k}^{-1}$, must have $q_{k} = 0$ in \Eref{Exp_Inv_QH}. Factors of $Z_{k}$ appear in $\prod_{i<j}\left(Z_{i} - Z_{j}\right)^{q}$ either from $\prod_{l(<k)}\left(Z_{l} - Z_{k}\right)^{q}$, or from $\prod_{l(>k)}\left(Z_{k} - Z_{l}\right)^{q}$. Note that both the binomial expansions of $\left(Z_{l} - Z_{k}\right)^{q}$ and $\left(Z_{k} - Z_{l}\right)^{q}$ produce homogeneous polynomials of degree $q$. The only monomial in these expansions with no powers of $Z_{k}$ is $Z_{l}^{q}$. This also shows that it is not possible to have two $q_{l}$ to be equal to zero.

Using this we write the singular (not normalized) part of the continuum quasielectron wavefunction \re{Laugh_1QE} as follows:
\bea 
\fl \big |\psi_{\mathrm{QE}}^{\infty}\big\rangle_{\mathrm{cont}}  = \nc^{-1} \int\cdots\int_{D^{\prime^{M}}} \prod_{i = 1}^{M}\Diff2 Z_{i} \big|\Zv\big\rangle \nonumber \\
\times \sum_{k = 1}^{M}Z_{k}^{-1}\prod_{l(<k)}Z_{l}^{q-1}\prod_{l(>k)}\left(-Z_{l}\right)^{q-1} \prod_{i<j \atop i,j\neq k} \left(Z_{i} - Z_{j}\right)^{q} \rme^{-\frac{|Z_{j}|^2}{4}},
\label{Laugh_1QE_Sing}
\eea   
where $\nc$ is the same as in \Esref{Laugh_1QE}, \re{Exp_Inv_QH}, and \re{Exp_Inv_QH_Reg}. After cancellation only one of these poles in $\prod_{l}Z_{l}^{-1}$ remain in the singular terms \re{Laugh_1QE_Sing}. The other poles coming from $\prod_{l(\neq k)}Z_{l}^{-1}$, in the term that has a residual pole $Z_{k}^{-1}$, reduce the power of $Z_{l}$ from $q$ to $q-1$.  Denoting the $\lbrace\sqv\rbrace$ with $q_{k} = 0$ as $\lbrace\ksqv\rbrace$ we finally write \Eref{Laugh_1QE_Sing} as follows:
\bea 
\fl \big |\psi_{\mathrm{QE}}^{\infty}\big\rangle_{\mathrm{cont}}  = \nc^{-1} \int\cdots\int_{D^{\prime^{M}}} \prod_{i = 1}^{M}\Diff2 Z_{i} 
 \big|\Zv\big\rangle \sum_{k = 1}^{M}Z_{k}^{-1} \nonumber \\ \times \sum_{\ksqv} \mathcal{E}_{\ksqv}\prod_{l (\neq k)}Z_{l}^{q^{k\infty}_{l}-1}\rme^{-\frac{|Z_{l}|^2}{4}}.
\label{Exp_Inv_QH_Sing}
\eea   

Using the orthogonality of the basis elements in the right hand side of \Eref{Exp_Inv_QH_SEB}, and \Esref{Exp_Inv_QH_Reg} and \re{Exp_Inv_QH_Sing}, we derive the following normalization condition for the wavefunction \re{Exp_Inv_QH}:
\beg
\nc^{2}_{\mathrm{reg}} + \nc^{2}_{\infty} = \nc^{2},
\label{Norm_Cond}
\en
where
\beg  
\fl \nc^{2}_{\mathrm{reg}} = \sum_{\rqv} \mathcal{E}^{2}_{\rqv}\prod_{l = 1}^{M}\int\int_{D^{\prime}} \Diff2 Z_{l} \left|Z_{l}\right|^{2\left(q^{\mathrm{reg}}_{l}-1\right)}\rme^{-\frac{|Z_{l}|^2}{2}}, 
\label{Norm_Cond_Reg}
\en
and 
\bea 
\fl \nc^{2}_{\infty} = \sum_{k = 1}^{M} \int\int_{D^{\prime}} \Diff2 Z_{k} \frac{1}{\left|Z_{k}\right|^{2}}\rme^{-\frac{|Z_{k}|^2}{2}} \nonumber \\ \times \sum_{\ksqv} \mathcal{E}^{2}_{\ksqv}\prod_{l (\neq k)}\int\int_{D^{\prime}} \Diff2 Z_{l}  \left|Z_{l}\right|^{2\left(q^{k\infty}_{l}-1\right)}\rme^{-\frac{|Z_{l}|^2}{2}}.
\label{Norm_Cond_Sing}
\eea
The integrands of the type $|Z_{l}|^{2\left(q_{l}-1\right)}\rme^{-\frac{|Z_{l}|^2}{2}}$ lead to the following radial integrals:
\beg
\int_{0}^{\infty} r_{l}^{2q_{l} - 1}\rme^{-r_{l}^2/2}dr = 2^{q_{l} - 1}\Gamma(q_{l}),
\label{typ_rad_Int}
\en
where in the $\xi \rightarrow 0$ limit we have simply replaced $\xi$ by zero. The integrands of the type $|Z_{k}|^{-2}\rme^{-\frac{|Z_{k}|^2}{2}}$ in \Eref{Norm_Cond_Sing}, on the other hand, lead to
\beg
\int_{\xi}^{\infty} \frac{1}{r}\rme^{-r^2/2}dr = \frac{1}{2}E_{1}\left(\frac{\xi^{2}}{2}\right),
\label{div_rad_Int}
\en
where $\xi \propto 1/\sqrt{N}$ is the distance of closest approach between the electrons and the singularity, and $E_{1}(x)$ is the exponential integral \cite{Abramowitz_Stegun}. To estimate the exponential integral we use the following bound:
\beg
\frac{1}{2}\rme^{-x}\ln{\left(1+\frac{2}{x}\right)}<E_{1}(x)<\rme^{-x}\ln{\left(1+\frac{1}{x}\right)},
\label{exp_Int}
\en
where $x>0$.

For comparing the contribution of the regular terms with the contribution of the irregular terms with poles, we introduce the following (always positive) ratio:
\beg
r = \frac{\nc^{2}_{\infty} }{\nc^{2}_{\mathrm{reg}}} = \frac{1}{2}E_{1}\left(\frac{\xi^{2}}{2}\right)r_{*}(M),
\label{Ratio_Div_Reg}
\en
where $r_{*}(M)$ denotes a function of $M$. In the above, we have used that each term of the sum $\sum_{k}$ in the right hand side of \Eref{Norm_Cond_Sing} is proportional to $E_{1}\left(\xi^{2}/2\right)/2$. From \Esref{exp_Int} and \re{Ratio_Div_Reg}, we show that in the continuum limit the contribution from the pole \re{Norm_Cond_Sing} diverges as $\ln N$, when the quasielectron is situated either at the center of a square plaquette or midway between two lattice sites. The slow nature of the singularity is indeed difficult to capture in numerics.

Consistent with the above analysis, we numerically demonstrate that in the divergence of the excess charge $\textbf{Q}_{-1}(r)$ the irregular terms (the ones leading to a pole in the continuum and hence not confined to the LLL) in ansatz \re{Latt_1QE} play the most important part. Here we are concerned about the continuous lowering of the $\textbf{Q}_{-1}(r)$ profile near the minima close to $r = 0.25\sqrt{2\pi}$. The other parts of the profile eventually converge. In particular, the oscillations for $r\gtrsim 2.0\sqrt{2\pi}$ due to the underlying lattice structure die off, see \Fref{Div_Terms_Responsible}\blue{a}. With the help of \Fsref{Div_Terms_Responsible}\blue{b}, \ref{Div_Terms_Responsible}\blue{c}, and \ref{Div_Terms_Responsible}\blue{d} we show that, when we decrease $\eta$, the $\textbf{Q}_{-1}(r)$ profile between $0<r\lesssim \sqrt{2\pi}$ is most affected by the aforementioned divergent terms.

\section{Properties of the LLL-Projected Quasielectron Ansatz}
\label{Sec: Failure_Div_Free}


\begin{figure}[tbp!]
\begin{indented}\item[]
\fl \raisebox{5cm}{\bf(a)}\label{eta_q_2X1X1_Comp}\includegraphics[trim={0.7cm 7cm 0.7cm 7cm},clip, scale=0.38]{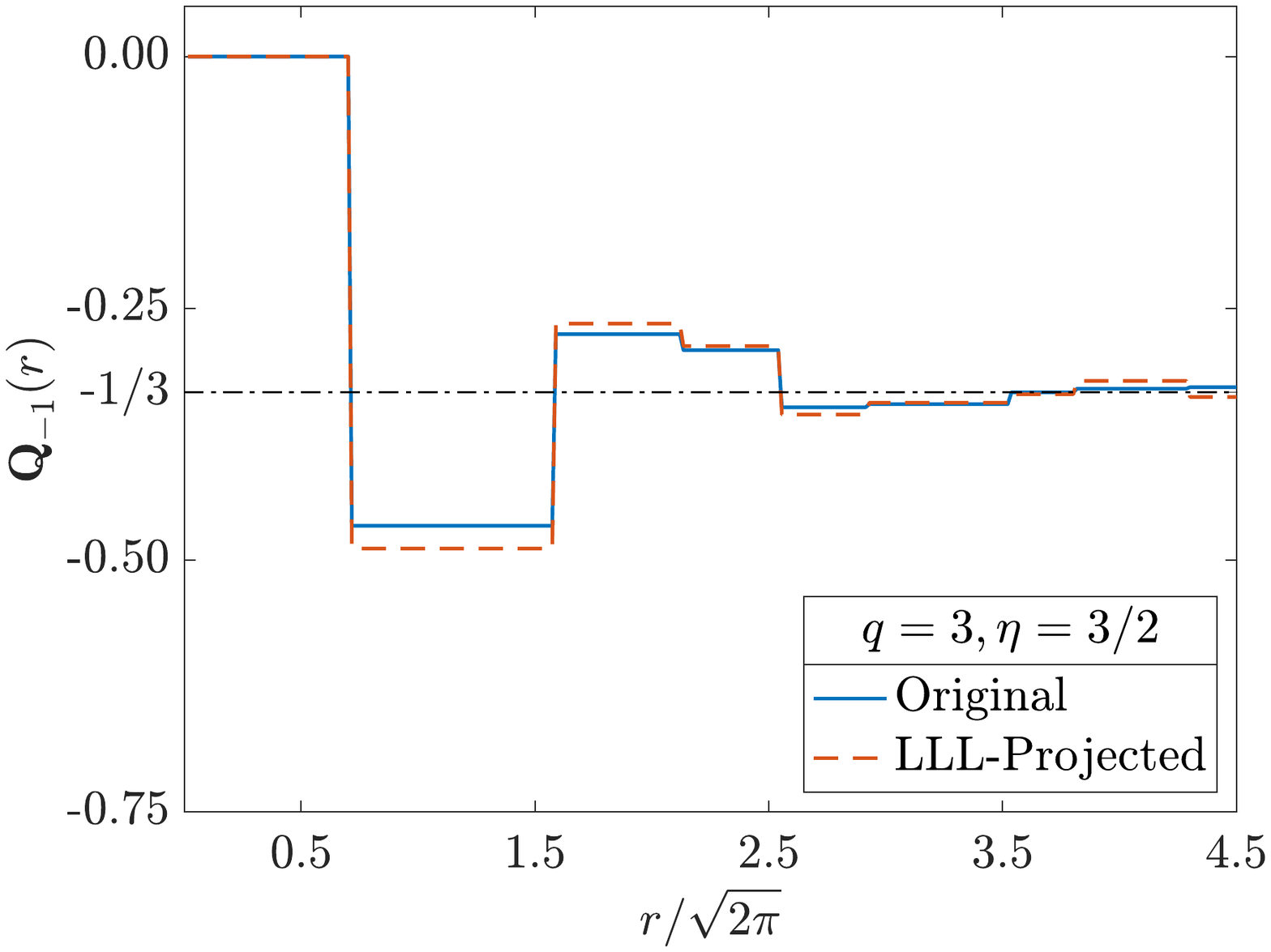}\quad
\raisebox{5cm}{\bf(b)} \includegraphics[trim={0.7cm 7cm 0.7cm 7cm},clip, scale=0.38]{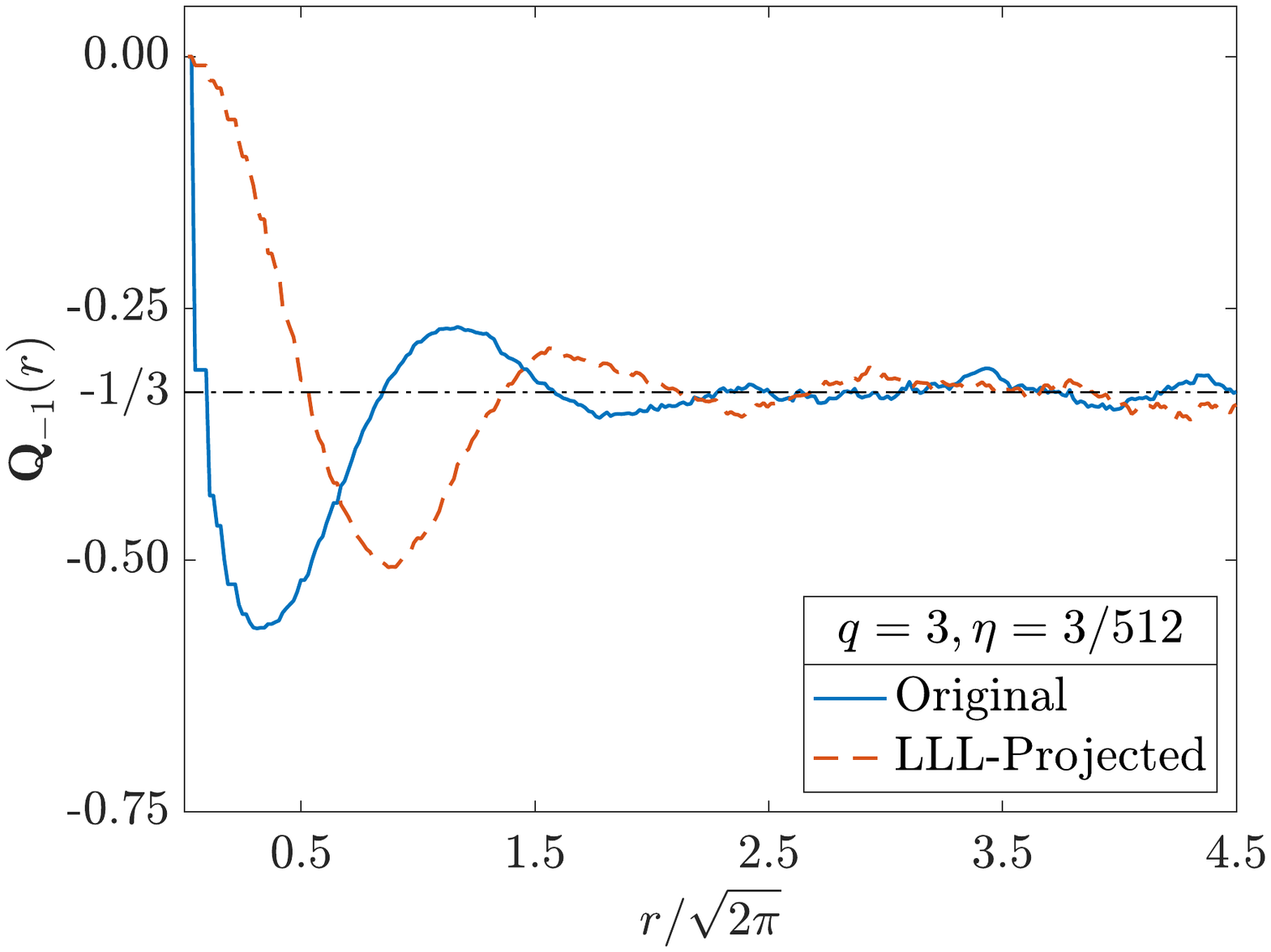}\llap{\raisebox{3.5cm}{\includegraphics[trim={0.5cm 7cm 1cm 7cm}, clip, height=2.6cm]{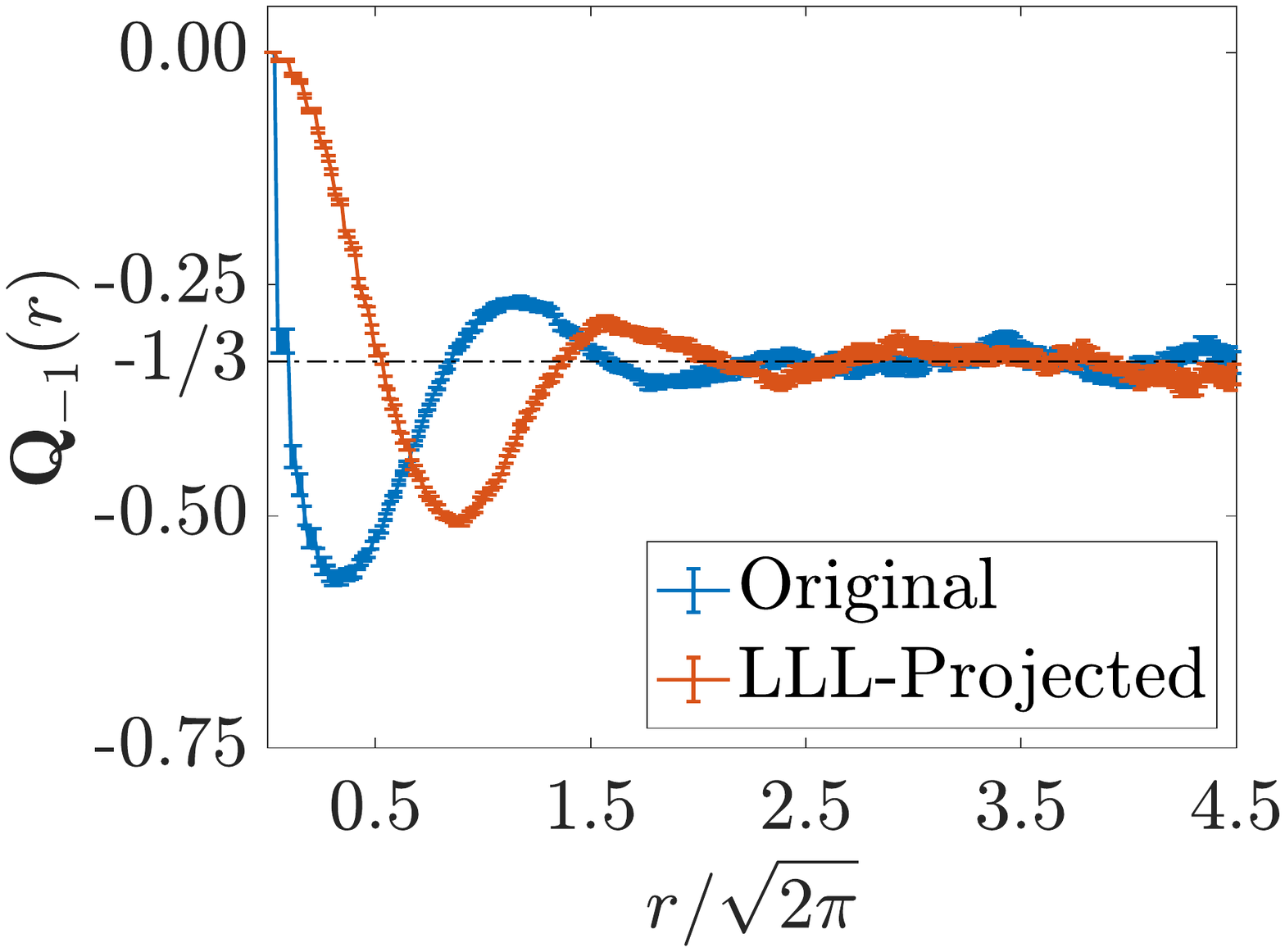}}}
\caption{Comparison of the excess charge $\textbf{Q}_{-1}(r)$ for the quasielectron, calculated with the original quasielectron ansatz \re{Latt_1QE} [blue solid] and the modified divergence-free ansatz \re{Latt_1QE_Mod} [red dashed]. Both of them are for a square lattice on a circular disk of radius $R \approx 8.3\sqrt{2\pi}$ and inverse filling fraction $q = 3$. We have $\eta = q/2$ in \textbf{(a)}, and $\eta = q/\left(2\times 16^{2}\right)$ in \textbf{(b)} [$\pm \sigma$ errorbar in the inset]. Already in \textbf{(a)} the excess charge profiles do not match fully. For large $r$, however, we observe that $\textbf{Q}_{-1}(r) \rightarrow 1/q$ for all the profiles.}
\label{Comarison_Orig_Corr}
\end{indented} 
\end{figure}


The obvious way to remove the singularity at the quasielectron position $w_{1}$ from the ansatz \re{Latt_1QE} is to project it to the LLL. We do this by removing the singular terms, which amounts to subtracting the following terms from $\prod_{i<j}\left(z_{i} - z_{j}\right)^{qn_{i}n_{j}}$:
\beg 
\mathcal{R} = \sum_{k = 1}^{N}n_{k}\prod_{l(>k)}\left(w_{1} - z_{l}\right)^{qn_{l}n_{k}} \prod_{l(<k)}\left(z_{l} - w_{1}\right)^{qn_{k}n_{l}} \prod_{i<j \atop i,j\neq k}\left(z_{i} - z_{j}\right)^{qn_{i}n_{j}}.
\label{QE_Subtracted_1}
\en 
Comparing \Eref{QE_Subtracted_1} with \Eref{Laugh_1QE_Sing}, we note that it is equivalent to subtracting the following from $\prod_{i<j}\left(Z_{i} - Z_{j}\right)^{q}$ in the continuum limit:
\beg 
\mathcal{R}_{\mathrm{cont}} = \sum_{k = 1}^{M}\prod_{l(>k)}\left(w_{1} - Z_{l}\right)^{q} \prod_{l(<k)}\left(Z_{l} - w_{1}\right)^{q} \prod_{i<j \atop i,j\neq k}\left(Z_{i} - Z_{j}\right)^{q}.
\label{QE_Subtracted_2}
\en 

Using this we write the modified lattice ansatz for a single quasielectron as follows:
\bea 
\fl \big|\psi_{\mathrm{QE}}\big\rangle = \mathcal{C}^{-1} \sum_{\nv}\delta_{n} \prod_{i = 1}^{N}\chi_{n_i}
\prod_{i = 1}^{N} \left(w_{1} - z_{i}\right)^{-n_{i}} \left(1 - \mathcal{F}\right) \nonumber \\ 
\times \prod_{i<j}\left(z_{i} - z_{j}\right)^{qn_{i}n_{j} - \eta\left(n_{i} + n_{j}\right)} \big |\nv \big\rangle,
\label{Latt_1QE_Mod}
\eea 
where we use phase factors $\chi_{n_i}$ similar to \Eref{Latt_1QE_1QH}, and use the following definition for $\mathcal{F}$:
\beg 
\mathcal{F} = \mathcal{R}\prod_{i<j}\left(z_{i} - z_{j}\right)^{-qn_{i}n_{j}}
= \sum_{k}n_{k}\prod_{l(\neq k)}\left(\frac{w_1 - z_{l}}{z_{k} - z_{l}}\right)^{qn_{l}n_{k}}.
\label{Mod_QE_Factor_2}
\en 
Incorporating the above modification, one writes the ansatz for a quasielectron at $w_{1}$ and a quasihole at $w_{2}$ as follows:
\bea 
\fl \big|\psi \big\rangle = \mathcal{C}^{-1} \sum_{\nv}\delta_{n} \prod_{i = 1}^{N}\chi_{n_i} \prod_{i = 1}^{N} \left(w_{1} - z_{i}\right)^{-n_{i}}\left(1 - \mathcal{F}\right) \prod_{i = 1}^{N}\left(w_{2} - z_{i}\right)^{n_{i}} \nonumber \\
\times \prod_{i<j}\left(z_{i} - z_{j}\right)^{qn_{i}n_{j} - \eta\left(n_{i} + n_{j}\right)} \big |\nv \big\rangle,
\label{Latt_1QE_1QH_Mod}
\eea 
where we have kept the multiplicative factor for the quasihole $\prod_{i = 1}^{N}\left(w_{2} - z_{i}\right)^{n_{i}}$ unchanged.

We numerically study the excess charge profiles with the modified divergence-free quasielectron \re{Latt_1QE_Mod} in \Fref{Comarison_Orig_Corr}. It is known that for $\eta = q/2$ the quasielectron excess charge is precisely the negative of the quasihole excess charge \cite{anqe}. For this special value of $\eta$ we see that the original quasielectron \re{Latt_1QE} and the modified divergence-free quasielectron \re{Latt_1QE_Mod} produce almost similar profiles for $\textbf{Q}_{-1}(r)$. However, as we move away from this special $\eta$ value and move toward the continuum limit, the difference between the two quasielectrons becomes more apparent. However, for large $r$ we observe $\textbf{Q}_{-1}(r) \rightarrow 1/q$ for both quasielectrons. We now demonstrate that the LLL-projected quasielectron does not have the properties expected for anyons hosted in systems with the topology of the Laughlin state.

\subsection{Inconsistencies in the Modified Ansatz}
\label{Sec: Incon_Mod_Ansatz}

In this section we first show that the modified ansatz does not give rise to the braiding properties \Eref{statistics}. We further demonstrate that starting from the ansatz for an FQH system with a quasihole and a modified quasielectron it is not possible to isolate a quasihole.

\subsubsection{Braiding Statistics}
\label{Sec: Incorr_Braid}

To determine the braiding statistics between a quasihole and the modified quasielectron, we need to obtain the Berry phases from the wavefunction \re{Latt_1QE_1QH_Mod}. We write the normalization constant there as follows:
\beg 
\fl \nc^{2} = \sum_{\nv}\delta_{n} \prod_{i}\left|w_{1} - z_{i}\right|^{-2n_{i}} \left|w_{2} - z_{i}\right|^{2n_{j}} \prod_{i<j}\left|z_{i} - z_{j}\right|^{2qn_{i}n_{j} - 2\eta\left(n_{i} + n_{j}\right)} \left|1 - \mathcal{F}\right|^{2}.
\en 

Following the same steps delineated in \Sref{Sec: Braiding_Lattice_Wavefunc}, we obtain the Berry phase by moving the quasihole coordinate $w_{2}$ in a loop as follows:
\beg
\theta_{2} = \frac{i}{2}\oint_{c}\sum_{i}\frac{\mean{n_{i}}}{w_{2} - z_{i}} dw_{2} + \mathrm{c.c.},
\label{BP_Exp_QH_Mod}
\en
where we have used the fact that $\mathcal{F}$ is $w_{2}$ independent. Using this expression for the Berry phase, and the fact that both the quasihole and modified quasielectron are localized, we obtain the anyonic statistics between the quasihole and the quasielectron to be $\gamma = -1/q$.

We then proceed to calculate the Berry phase by moving the quasielectron coordinate $w_1$ in a loop and obtain
\beg 
\theta_{1} = -\frac{i}{2}\oint_{c}\sum_{i}\frac{\mean{n_{i}}}{w_{1} - z_{i}} dw_{1}  - \frac{i}{2}\oint_{c}\left\langle\frac{1}{1 - \mathcal{F}}\frac{\partial \mathcal{F}}{\partial w_{1}}\right\rangle dw_{1} + \mathrm{c.c.}.
\label{Correct_BP_QE_Move}
\en 
This Berry phase does not give the braiding statistics expected for anyons because of the additional term. Moreover, this additional term contains nontrivial poles of $w_{1}$ that are functions of the lattice positions $z_{i}$. As a result, it depends on the size and shape of the contour $c$.

\subsubsection{Isolating a Quasihole}
\label{Sec: Isolated_QH}

We explained above \Eref{Latt_1QE}, how taking the limit $w_{2}\rightarrow \infty$ in \Eref{Latt_1QE_1QH} leads to an isolated quasielectron at $w_{1}$. One can similarly take the $w_{1}\rightarrow \infty$ limit. In \Eref{Latt_1QE_1QH}, $\prod_{i = 1}^{N}\left(w_{1} - z_{i}\right)^{-n_{i}}$ becomes a $z_{i}$ independent factor $w^{-M}_{1}$ in that limit, which is absorbed in the normalization. Evidently this limit isolates the quasihole at $w_{2}$.

We take the limit $w_{1} \rightarrow \infty$ in \Eref{Latt_1QE_1QH_Mod}. We only need to investigate the behavior of $\left[1 - \mathcal{F}\right]$. Since $w_{1}$ is in the numerator of $\mathcal{F}$, we neglect the one in front. The problem is that this factor contains factors like $\left(z_{l} - z_{k}\right)^{qn_{l}n_{k}}$. Because of such dependence on $\zv$, we cannot absorb $\mathcal{F}$ in the normalization.

\section{Conclusion}
\label{Sec: Conclusion}

In this paper, we studied the continuum limit of a lattice quasielectron ansatz. This lattice quasielectron, besides having the topological properties expected for an anyonic quasielectron, is the inverse of the lattice quasihole.

Our study reveals that a continuum limit of this lattice wavefunction does not exist. Only when the quasielectron is placed on top of a lattice site, this limit of the lattice quasielectron gives a finite result. Otherwise, depending on how we approach it, in the continuum limit the quasielectron pole gives rise to a slow $\ln N$ singularity, where $N$ is the number of lattice sites. This also explains why such divergence is hard to observe in the numerics. We then try to salvage the ansatz wavefunction by projecting it on the LLL, which removes the singularity. We demonstrate that the projected state does not obtain the braiding properties expected for anyons. We hence conclude that taking the continuum limit of the lattice quasielectron wavefunction does not lead to a suitable ansatz for anyonic quasielectrons in the continuum.

\ack
We thank Maria Hermanns and Eddy Ardonne for several helpful discussions and comments on the manuscript.



\end{document}